\documentclass{tlpbis}

\jnlPage{xx}{yy}
\jnlDoiYr{2024}
\doival{10.1017/xxxxx}

\usepackage{url}
\usepackage{booktabs}
\usepackage{graphicx}
\usepackage{stmaryrd}
\usepackage{xcolor}
\usepackage{xargs}
\usepackage{textcomp}
\usepackage{amsmath}
\usepackage{amssymb}
\usepackage{relsize}
\usepackage{xspace}
\usepackage{amsfonts}
\usepackage{mathtools}
\usepackage{framed}
\usepackage{subfig}
\usepackage{algorithmic}
\usepackage{multirow}
\usepackage{multicol}
\usepackage{array}
\usepackage{longtable}
\usepackage{makecell}
\usepackage[hidelinks]{hyperref}
\usepackage[depth=4]{bookmark}
\usepackage{patchcmd}

\newcommand{\figpath}{./Figs}

\newtheorem{example}{Example} %

\DeclareMathOperator*{\argmin}{arg\,min}

\newcommand{\lbr}{\mbox{$\lbrack\hspace{-0.3ex}\lbrack$}}
\newcommand{\rbr}{\mbox{$\rbrack\hspace{-0.3ex}\rbrack$}}
\newcommand{\eval}[2]{\lbr #1 \rbr_{#2}}

\newcommand{\good}{$\mathbf{\checkmark}$}
\newcommand{\approxbnd}{$\Theta$}
\newcommand{\approxnobnd}{$\Theta$}
\newcommand{\bad}{$\--$}
\newcommand{\none}{$\cdot$}

\newcommand{\norm}[1]{\left\lVert#1\right\rVert}

\usepackage[colorinlistoftodos,textwidth=2.2cm]{todonotes}

\usepackage{tikz}
\usetikzlibrary{arrows.meta,positioning,shapes,shadows}

\usepackage{verbatim}
\usepackage[titlenumbered,ruled,linesnumbered]{algorithm2e}
\usepackage{listings}
\newcommand{\prettylstciao}[0]{
\lstset{language=Prolog,
  frameround=tttt,
  xleftmargin=0.2cm,
  rulecolor=\color{blue},
  numbers=left,numberstyle=\tiny,stepnumber=1,numbersep=8pt,
  tabsize=4,
  showstringspaces=false,
  breaklines=true,breakatwhitespace=true,
  showlines=true,
  showspaces=false,showtabs=false,%
  upquote=true,
  commentstyle=\color{gray},
  keywordstyle=\color{eminence},
  basicstyle=\small\ttfamily,
  keywordstyle=\color{weborange},
  emphstyle={\color{blue}},
  emph={pred,prop,trust,check,checked,true,rsize,cardinality,
  not_fails,module,exp,cost,steps_ub,steps_lb,size_ub,size_lb,
  covered,mut_exclusive,cost,use_module,int,calls,success,cost_center,
  is_det,num,nat,var,list,ground,length,terminates,term,steps_o,resource,
  entry,impl_defined,regtype,tick},
  otherkeywords={>,<,>=,=<,.,;,-,!,=,~,*,\&,+,:-,[,],|,->,:,:=},
  morekeywords= {>,<,>=,=<,.,;,-,!,=,~,*,\&,+,:-,[,],|,->,:,:=},
  escapeinside=\`\`,
}}

\newcommand{\ciao}{Ciao\xspace}
\newcommand{\ciaopp}{CiaoPP\xspace}

\definecolor{eminence}{RGB}{108,48,130}
\definecolor{weborange}{RGB}{200,130,0}

\newcommand\stdub[2]{{\tt C}_{#1}(#2)}
\newcommand\sizeub[2]{{\tt S}_{#1}(#2)}

\newcommand\costrelations{cost relations}

\newcommand{\recf}{f} %
\newcommand{\numparam}{m} %
\newcommand{\candf}{\hat{f}} %
\newcommand{\basefuns}{\mathcal{F}} %
\newcommand{\score}{\mathcal{S}} %
\newcommand{\numfeatures}{p} %
\newcommand{\trainingset}{\mathcal{T}}
\newcommand{\numtraining}{n} %
\newcommand{\bndtraining}{b} %
\newcommand{\inputset}{\mathcal{I}} %
\newcommand{\sequence}{S}
\newcommand{\realnumbers}{\mathbb{R}}
\newcommand{\integernumbers}{\mathbb{Z}}
\newcommand{\naturalnumbers}{\mathbb{N}}
\newcommand{\rr}{\mathbb{R}}
\newcommand{\qq}{\mathbb{Q}}
\newcommand{\nn}{\mathbb{N}}
\newcommand{\dd}{\mathcal{D}}
\newcommand{\pp}{\mathcal{P}}
\newcommand{\evalfun}{E}
\newcommand{\selcomb}{\psi} 
  
\newcommand{\tuple}[1]{\langle #1 \rangle}
\newcommand\vvec[1]{\vec{\vphantom{l}\smash{#1}}} %

\begin{document}

\pgfdeclarelayer{marx}
\pgfsetlayers{main,marx}
\providecommand{\cmark}[2][]{%
  \begin{pgfonlayer}{marx}
    \node [nmark] at (c#2#1) {#2};
  \end{pgfonlayer}{marx}
  }
\providecommand{\cmark}[2][]{\relax}

\lefttitle{L.~Rustenholz, M.~Klemen, M.A.~Carreira-Perpi\~{n}\'{a}n, P.~Lopez-Garcia}

\title[Machine Learning-based Recurrence Relation Solving]{A Machine
  Learning-based Approach for Solving Recurrence Relations and its use
  in Cost Analysis of Logic Programs
\footnote[2]{Extended,
revised version
of our work published in ICLP~\citep{ml-rec-solving-iclp2023} (see~\ref{appendix:new-contributions}).}
\thanks{
Research
partially supported by MICINN projects
PID2019-108528RB-C21 \emph{ProCode}, TED2021-132464B-I00
\emph{PRODIGY}, and FJC2021-047102-I,
and the Tezos foundation.  The authors would also like to thank the
anonymous reviewers for their insightful comments, which really help
improve this paper. Special thanks go to ThanhVu Nguyen for his
assistance with the related work discussion, and to John Gallagher,
Manuel V.  Hermenegildo and Jos\'{e} F. Morales for
their valuable discussions, feedback, and their work as developers of
the \ciaopp system.}}

\begin{authgrp}
\author{\gn{Louis} \sn{Rustenholz}, \gn{Maximiliano} \sn{Klemen}}
\affiliation{Technical University of Madrid (UPM) \& IMDEA Software Institute}
\author{\gn{Miguel \'{A}.} \sn{Carreira-Perpi\~{n}\'{a}n}}
\affiliation{University of California, Merced}
\author{\gn{Pedro} \sn{Lopez-Garcia}}
\affiliation{Spanish Council for Scientific Research \& IMDEA Software Institute}
\end{authgrp}

\maketitle

\begin{abstract}
Automatic static cost analysis infers information about the resources
used by programs without actually running them with concrete data, and
presents such information as functions of input data sizes.  Most of
the analysis tools for logic programs (and many for other languages),
as \ciaopp,
are based on setting up recurrence relations representing
(bounds on)
the computational cost of predicates and solving them to find
closed-form functions. 
Such recurrence solving is a bottleneck in current tools: many
of the recurrences that arise during the analysis cannot be solved
with state-of-the-art
solvers,
including Computer Algebra Systems (CASs),
so that specific methods for
different classes of recurrences need to be developed.  We address
such a challenge by developing a novel, general approach for solving
arbitrary, constrained recurrence relations, that uses
machine learning (sparse-linear and symbolic) regression techniques to
\emph{guess} a candidate closed-form function, and a combination of an
SMT-solver and a CAS to \emph{check} whether such function is actually
a solution of the recurrence.
Our prototype implementation and its experimental evaluation within
the context of the \ciaopp system
show quite promising results.
Overall, for the considered benchmark set, our approach outperforms
state-of-the-art cost analyzers and recurrence solvers, and can find
closed-form solutions, in a reasonable time, for recurrences that
cannot be solved by them.

\phantom{.}

Under consideration in Theory and Practice of Logic Programming (TPLP).
\end{abstract}


\section{Introduction and Motivation}
\label{sec:intro}

The motivation of the work presented in this paper stems from
automatic static cost analysis and verification of logic
programs~\citep{granularity-short,caslog-short,low-bounds-ilps97-short,resource-iclp07-short,plai-resources-iclp14-short,gen-staticprofiling-iclp16-short,resource-verification-tplp18-shortest}.
The goal of
such analysis is to infer information about the resources used by
programs without actually running them with concrete data, and present
such information as functions of input data sizes and possibly other
(environmental) parameters.
We assume a broad concept of resource as a numerical property of the
execution of a program, such as number of \emph{resolution steps},
\emph{execution time}, \emph{energy consumption}, \emph{memory},
number of \emph{calls} to a predicate,
number of \emph{transactions} in a database, etc.
Estimating in advance the resource usage of computations is useful for
a number of applications, such as
automatic program optimization,
verification of resource-related specifications, detection of
performance bugs, helping developers make resource-related design
decisions,
security applications (e.g., detection of side
channel attacks),
or blockchain platforms (e.g., smart-contract gas analysis and
verification).

The challenge we address originates from the established approach of
setting up recurrence relations representing the cost of predicates,
parameterized by input data
sizes~\citep{Wegbreit75,Rosendahl89,granularity-short,caslog-short,low-bounds-ilps97-short,resource-iclp07-short,AlbertAGP11a-short,plai-resources-iclp14-short,gen-staticprofiling-iclp16-short},
which are then solved to obtain \emph{closed forms} of such
recurrences (i.e., functions that provide either exact, or upper/lower
bounds on resource usage in general).
Such approach can infer different classes of functions (e.g.,
polynomial, factorial, exponential, summation, or logarithmic).

The applicability of these resource analysis techniques strongly
depends on the capabilities of the component in charge of solving (or
safely approximating) the recurrence relations generated during the
analysis, which has become a bottleneck in some systems.

A common approach to automatically solving such
recurrence relations
consists of
using a Computer Algebra System (CAS) or a specialized solver.
However, this approach poses several difficulties and
limitations.
For example,
some recurrence relations contain complex expressions or recursive
structures that most of the well-known CASs cannot solve, making
it necessary to develop ad-hoc techniques to handle such cases.
Moreover, some recurrences may not have the form required by such
systems because an input data size variable does not decrease, but
increases instead. Note that a decreasing-size variable could be
implicit in the program, i.e., it could
be a function of a subset of input data sizes (a ranking function), which
could be inferred by applying established techniques
used in termination analysis~\citep{PodelskiR04b-short}. However, such
techniques are usually restricted to linear arithmetic.

In order to address this challenge
we have developed a novel, general method for solving arbitrary,
constrained recurrence relations.
It is a \emph{guess and check} approach that uses
machine learning techniques for the \emph{guess} stage, and a
combination of an SMT-solver and a CAS for the
\emph{check} stage (see Figure~\ref{fig:mlresol-arch}).
To the best of our knowledge, there is no
other approach that does this.
The resulting closed-form function solutions can be of different
kinds, such as polynomial, factorial, exponential or logarithmic.

Our method is parametric in the \emph{guess} procedure used, providing
flexibility in the kind of functions that can be inferred, trading off
expressivity with efficiency and theoretical guarantees.
We present the results of two instantiations, based on sparse linear
regression (with non-linear templates) and symbolic regression.
Solutions to our equations are first evaluated on finite samples of
the input space, and then, the chosen regression method is applied.
In addition to obtaining exact solutions, the search and optimization
algorithms typical of machine learning methods enable the efficient
discovery of good approximate solutions in situations where exact
solutions are too complex to find. These approximate solutions are
particularly valuable in certain applications of cost analysis, such
as granularity control in parallel/distributed computing.
Furthermore, these methods allow exploration of model spaces that
cannot be handled by complete methods such as classical linear
equation solving.

\begin{figure*}[t]
\centering
\resizebox{1\textwidth}{!}{
\centering 
\pgfdeclarelayer{background}
\pgfdeclarelayer{foreground}
\pgfsetlayers{background,main,foreground}


\tikzstyle{input}  = [
  chamfered rectangle, chamfered rectangle corners=north west,
  draw=cyan!80!black!100, fill=cyan!20,
  minimum width=2cm, minimum height=1cm,
  text centered, font=\sffamily
]

\tikzstyle{io}  = [
  chamfered rectangle, chamfered rectangle corners=north west,
  draw=red!80!black!100, fill=red!30,
  minimum width=2cm, minimum height=2cm,
  text centered, font=\sffamily
]

\tikzstyle{output}  = [
  chamfered rectangle, chamfered rectangle corners = north west,
  draw=cyan!80!black!100, fill=cyan!20, 
  minimum width = 3cm, minimum height = 3cm, 
  text centered, font=\sffamily
]

\tikzstyle{approxoutput}  = [
  chamfered rectangle, chamfered rectangle corners=north west,
  draw=red!80!black!100, fill=red!30,
  minimum width=3cm, minimum height=3cm,
  text centered, font=\sffamily
]

\tikzstyle{process}=[
  rectangle, rounded corners,  
  draw=green!50!black!100, fill=green!40,
  minimum width=3cm, minimum height=4cm, 
  thick,
  font=\sffamily
]

\tikzstyle{component}=[
  rectangle, rounded corners,  
  draw=orange!50!black!100, fill=orange!40,
  minimum width=2.5cm, minimum height=0.5cm, 
  thick,
  text centered, font=\sffamily
]

\tikzstyle{decision}=[
  diamond,
  draw=green!50!black!100, fill=green!40,
  minimum width=0.5cm, minimum height=0.5cm, 
  text centered, font=\sffamily
]

\tikzstyle{arrow} = [thick,->]

\begin{tikzpicture}
  [every node/.style={align=center}, >=latex, scale=1]

\node (recurrence) [input, xshift=-1cm] {\textbf{Recurrence} \\ [1mm] \textbf{Relation}};

\node (guesser) [process, right=0.7cm of recurrence] { } ;

\path [color=black, font=\sffamily] (guesser.north)+(0,-3em) node (guesser-tex)
      {\large{\textbf{Guesser}} \\ [1mm] Machine Learning};

\node (lasso) [component, below=0.3cm of guesser-tex] {\textbf{Lasso Regr.}};

\node (symb-regr) [component, below=0.3cm of lasso] {\textbf{Symb. Regr.}};

\node (candidate) [io, right=0.5cm of guesser]
      {\textbf{Candidate} \\ [1mm] \textbf{solution} \\ [1mm] \textbf{+} \\ [1mm] \textbf{Accuracy score}};

\node (checker) [process, right=0.5cm of candidate] { };

\path [color=black, font=\sffamily] (checker.north)+(0,-1.5em) node (checker-tex) {\large \textbf{Checker}};

\node (smt) [component, below=0.6cm of checker-tex] {\textbf{SMT Solver}};

\node (cas) [component, below=0.3cm of smt] {\textbf{CAS}};

\node (ifchecked) [decision, right=0.5cm of checker] {\textbf{Checked?} };

\node (solution) [output, right=0.7cm of ifchecked, yshift=+2cm]
      {\textbf{Exact} \\ [1mm] \textbf{closed-form} \\ [1mm] \textbf{solution}};
      
\node (nosolution) [approxoutput, right=0.7cm of ifchecked, yshift=-1.9cm]
      {\textbf{Approximate} \\ [1mm] \textbf{closed-form} \\ [1mm] \textbf{+} \\ [1mm]
       \textbf{Accuracy score} \\ [1mm] \textbf{(+)} \\ [1mm] \textbf{Counterexample}};

\path[arrow]  (recurrence) edge (guesser)      
              (guesser)    edge (candidate)
              (candidate)  edge (checker)
              (checker)    edge (ifchecked);
\draw[arrow] (ifchecked)  |- node[pos=0.25, right] {\textbf{yes}} (solution);
\draw[arrow] (ifchecked)  |- node[pos=0.25, right] {\textbf{no}} (nosolution);

\begin{pgfonlayer}{background}
  \path (guesser.north west)+(-0.3,1.5) node (g) {};
  \path (ifchecked.south east)+(0.8,-3.0) node (h) {};
  \path[fill=yellow!20,rounded corners, draw=black!50, densely dashed] (g) rectangle (h);
\end{pgfonlayer}

\end{tikzpicture}
}
\caption{Architecture of our novel machine learning-based recurrence solver.}
\label{fig:mlresol-arch}
\end{figure*}

The rest of this paper is organized as follows.
Section~\ref{sec:overview} gives and overview of our novel \emph{guess
and check} approach. Then, Section~\ref{sec:prelim} provides some
background information and preliminary notation.
Section~\ref{sec:description-approach} presents a more detailed,
formal and algorithmic description of our approach.
Section~\ref{sec:application-cost-analysis} describes the use of our
approach in the context of static cost analysis of (logic) programs.
Section~\ref{sec:implem-and-eval} comments on our prototype
implementation as well as its experimental evaluation and comparison
with other solvers.
Finally, Section~\ref{sec:related-work} discusses related work,
and Section~\ref{sec:conclusions} summarizes some conclusions and
lines for future work.
Additionally, \ref{appendix:full-output} provides complementary data
to the evaluation of Section~\ref{sec:implem-and-eval} and
Table~\ref{table:newexp1}.

\section{Overview of our Approach}
\label{sec:overview}

We now give an overview of our approach and its two stages already
mentioned, illustrated in Figure~\ref{fig:mlresol-arch}: \emph{guess}
a candidate closed-form function, and \emph{check} whether such
function is actually a solution of the recurrence relation.

Given a recurrence relation for a function $f(\vec{x})$, solving it means to
find a closed-form expression $\candf(\vec{x})$ defining a function, on the
appropriate domain, that satisfies the relation. We say that $\candf$ is a
closed-form expression whenever it does not contain any subexpression built
using $\candf$ (i.e., $\candf$ is not recursively defined), although we will
often additionally aim for expressions built only on elementary arithmetic
functions, e.g., constants, addition, subtraction, multiplication, division,
exponential, or perhaps rounding operators and factorial.
We will use the following recurrence as an example to illustrate our approach.
\begin{equation}
\label{eq:recurrence}
  \begin{array}{ll}
    f(x) \, = \; 0 &  \text{ if } x = 0 \\ %
    f(x) \, = \; f(f(x-1)) + 1 &  \text{ if } x > 0 
  \end{array}
\end{equation}
\noindent

\subsection{The ``guess'' stage}
As already stated, our method is parametric in the guess procedure
utilized, and we instantiate it with both sparse linear and symbolic
regression in our experiments. However, for the sake of presentation,
we initially focus on the former to provide an overview of our
approach. Subsequently, we describe the latter in
Section~\ref{sec:prelim}.  Accordingly, any possible model we can
obtain through sparse linear regression (which constitutes a candidate
solution) must be an affine combination of a predefined set of terms
that we call \emph{base functions}.
In addition,
we aim to use only a small number of such \emph{base functions}.
That is, a candidate solution is a function $\candf(\vec{x})$ of the
form
\begin{equation*}
 \candf(\vec{x}) = \beta_0 + \beta_1 \ t_1(\vec{x}) + \beta_2
 \ t_2(\vec{x}) + \ldots + \beta_{\numfeatures}
 \ t_{\numfeatures}(\vec{x}),
\end{equation*}
\noindent
where the $t_i$ are arbitrary functions on $\vec{x}$ from a set
$\basefuns$ of candidate
\emph{base functions},
which are representative of common complexity orders, and the
$\beta_i$'s are the coefficients (real numbers) that are estimated by
regression, but so that only a few coefficients are nonzero.

For illustration purposes, assume that we
use the following set $\basefuns$ of
base functions:
\begin{equation*}
    \begin{array}{rll} \basefuns = \{ \lambda x.x, \lambda x.x^2,
      \lambda x.x^3, \lambda x.\lceil \log_2(x) \rceil, \lambda x. 2^x,
\lambda x.x \cdot \lceil \log_2(x) \rceil \},
    \end{array}
\end{equation*}
\noindent
where each base function is represented as a lambda expression.  Then,
the sparse linear regression is performed as
follows.
\begin{enumerate}
\item Generate a training set $\trainingset$.
First, a set $\inputset = \{ \vec{x}_1, \ldots,\vec{x}_{\numtraining} \}$ of input
values to the recurrence function is randomly generated.  Then,
starting with an initial $\trainingset = \emptyset$, for each input value
$\vec{x}_i \in \inputset$, a training case $s_i$ is generated and
added to $\trainingset$.
For any input value $\vec{x} \in \inputset$ the corresponding training
case $s$ is a
pair of the form
\begin{equation*}
s = \big(\tuple{b_1, \ldots, b_{\numfeatures}},\,r\big),
\end{equation*}
\noindent
where $b_i = \eval{t_i}{\vec{x}}$ for $1 \leq i \leq \numfeatures$, and
$\eval{t_i}{\vec{x}}$ represents the result (a scalar) of evaluating
the base function $t_i \in \basefuns$ for input value $\vec{x}$, where $\basefuns$ is
a set of $\numfeatures$ base functions, as already explained. The (dependent)
value $r$ (also a constant number) is the result of evaluating the
recurrence $f(\vec{x})$ that we want to solve or approximate, in our
example, the one defined in Equation~\ref{eq:recurrence}.
Assuming that there is an $\vec{x} \in \inputset$ such that $\vec{x} =
\tuple{5}$, its corresponding training case $s$ in our example
will be
\begin{equation*}
   \begin{array}{rll}
     s &=& \big(\langle
     \eval{x}{5}, \eval{x^2}{5}, \eval{x^3}{5}, \eval{\lceil \log_2(x) \rceil}{5},
     \ldots \rangle,\,
     \mathbf{f(5)}\big) \\ &=& \big(\langle 5, 25, 125, 3, \ldots
     \rangle,\, \mathbf{5}\big). \\
   \end{array}
\end{equation*}
\item Perform sparse regression
  using the training
  set $\trainingset$ created above in order to find a small subset of base functions that fits it well.

  We do this in two steps. First, we solve an $\ell_1$-regularized linear
  regression to learn an estimate of the non-zero coefficients of the base
  functions. This procedure, also called lasso \citep{Hastie-15a}, was
  originally introduced to learn interpretable models by selecting a subset of
  the input features. This happens because the $\ell_1$ (sum of absolute
  values) penalty results in some coefficients becoming exactly zero (unlike
  an $\ell^2_2$ penalty, which penalizes the magnitudes of the coefficients
  but typically results in none of them being zero). This will typically
  discard most of the base functions in $\basefuns$, and only those that are really
  needed to approximate our target function will be kept. The level of
  penalization is controlled by a hyperparameter $\lambda \ge 0$. As commonly
  done in machine learning~\citep{Hastie-15a}, the value of $\lambda$ that
  generalizes optimally on unseen (test) inputs is found via cross-validation on
  a separate validation set (generated randomly in the same way as the training
  set).
  
  The result of this first sparse regression step are coefficients
  $\beta_1,\dots,\beta_{\numfeatures}$ (typically many of which are zero), and an
  independent coefficient $\beta_0$. In a second step, we keep only those
  coefficients (and their corresponding terms $t_i$) for which
  $|\beta_i|\geq\epsilon$ (where the value of $\epsilon \ge 0$ is determined
  experimentally). We find that this post-processing results in solutions that
  better estimate the true non-zero coefficients.
\item Finally, our method
  performs again a standard linear regression (without $\ell_1$
  regularization) on the training set $\trainingset$, but using only the base functions
  selected in the previous step.
  In our example, with $\epsilon = 0.05$, we obtain the model
  \begin{equation*}
    \begin{array}{rl}
      \candf(x) = 1.0 \cdot x. &
    \end{array}
  \end{equation*}
\end{enumerate}

A final test set $\trainingset_{\text{test}}$ with input set $\inputset_{\text{test}}$
is then generated (in the same way as the training set) to obtain a measure
$R^2$ of the accuracy of the estimation. In this case, we obtain a value $R^2 =
1$, which means that the estimation obtained predicts exactly the values for the
test set. This does not prove that the $\candf$ is a solution of the
recurrence, but this makes it a candidate solution for verification. If $R^2$
were less than $1$, it would mean that the function obtained is not a candidate
(exact) solution, but
an approximation (not necessarily a bound), as there are values in
the test set that cannot be exactly predicted.

Currently, the set of base functions $\basefuns$ is fixed;
nevertheless, we plan to automatically infer better, more
problem-oriented sets by using different heuristics, as we comment on
in Section~\ref{sec:conclusions}. Alternatively, as already mentioned,
our guessing method is parametric and can also be instantiated to
symbolic regression, which mitigates this limitation by creating new
expressions, i.e., new ``templates'', beyond linear combinations of
$\basefuns$. However, only shallow expressions are reachable in
acceptable time by symbolic regression's exploration strategy:
timeouts will often occur if solutions can only be expressed by deep
expressions.

\subsection{The ``check'' stage}

Once a function that is a candidate solution for the recurrence has
been guessed, the second step of our method
tries to verify whether such a candidate is actually a solution. To do
so,
the recurrence
is encoded as a first order
logic formula
where the references to the target function are replaced by the
candidate solution whenever possible. Afterwards, we use an
SMT-solver to check whether
the negation of such formula is satisfiable, in which case we can
conclude that the candidate is not a solution for the
recurrence. Otherwise, if such formula is unsatisfiable, then the
candidate function is
an exact solution. Sometimes, it is necessary
to consider a precondition for the domain of the recurrence, which is
also included in the encoding.

To illustrate this process,
Expression~\ref{eq:verexample1-rec} below 
\begin{equation}
  \label{eq:verexample1-rec}
  \begin{array}{ll}
  f(x) = \begin{cases*}
   0             & if $x = 0$ \\[-2pt]
   f(f(x-1)) + 1 & if $x > 0$
  \end{cases*}
  \end{array}
\end{equation}
shows the recurrence
we target to solve, for which the candidate solution obtained
previously using (sparse) linear regression is $\candf(x) = x$ (for
all $x\geq 0$).
Now, Expression~\ref{eq:verexample1-form} below shows the encoding
of the recurrence as a first order logic formula.
\begin{equation}
  \label{eq:verexample1-form}
  \vspace{-2pt}
  \forall x \: \Big( (x = 0 \implies \underline{f(x)} = 0)  \wedge (x>0 \implies \underline{f(x)} =
  \underline{f(\underline{f(x-1)})} + 1) \Big)
\end{equation}
\noindent
Finally, Expression~\ref{eq:verexample1-negform} below shows the negation
of such formula, as well as the references to the function name
substituted by the definition of the candidate solution. We underline
both the subexpressions to be replaced, and the subexpressions
resulting from the substitutions.
\begin{equation}
  \label{eq:verexample1-negform}
  \exists x \: \neg (\left( (x = 0 \implies \underline{x} = 0)  \wedge (x>0 \implies \underline{x} = \underline{x-1} + 1) \right))
\end{equation}
\noindent
It is easy to see that Formula~(\ref{eq:verexample1-negform})
is unsatisfiable. Therefore, $\candf(x) = x$ is an
exact solution for $f(x)$ in the recurrence defined by
Equation~\ref{eq:recurrence}.

For some cases where the candidate solution contains transcendental
functions, our
implementation of the method uses
a CAS to perform simplifications and
transformations, in order to obtain a formula supported by
the SMT-solver. We find this combination of CAS and SMT-solver
particularly useful, since it allows us to solve more problems than only
using one of these systems in isolation.

\section{Preliminaries}
\label{sec:prelim}

\paragraph[Notations]{\textbf{Notations}.}

We use the letters $x$, $y$, $z$ to denote variables, and $a$, $b$,
$c$, $d$ to denote constants and coefficients.  We use $f, g$ to
represent functions, and $e, t$ to represent arbitrary expressions. We
use $\varphi$ to represent arbitrary boolean constraints over a set of
variables. Sometimes, we also use $\beta$ to represent coefficients
obtained with (sparse) linear regression. In all cases, the symbols can be
subscribed.
$\nn$ and $\realnumbers^+$ denote the sets of non-negative integer and
non-negative real numbers respectively, both including $0$.
Given two sets $A$ and $B$, $B^A$ is the set of all functions from $A$
to $B$.
We use $\vec{x}$ to denote a finite sequence $\langle
x_1,x_2,\ldots,x_{\numfeatures}\rangle$, for some $\numfeatures>0$. Given a sequence $\sequence$ and
an element $x$, $\langle x| \sequence\rangle$ is a new sequence with first
element $x$ and tail $\sequence$.
We refer to the classical finite-dimensional $1$-norm (Manhattan norm) and
$2$-norm (Euclidean norm) by $\ell_1$ and $\ell_2$ respectively, while
$\ell_0$ denotes the ``norm'' (which we will call a pseudo-norm) which
counts the number of non-zero coordinates of a vector.

\paragraph[Recurrence relations]{\textbf{Recurrence relations}.}

In our setting, a \emph{recurrence relation} (or \emph{recurrence equation}) is
just
a functional equation on $\recf:\dd\to\rr$ with $\dd \subset \nn^{\numparam}$, $\numparam \geq
1$, that can be written as%
$$\forall \vec{x}\in\dd,\;\recf(\vec{x}) = \Phi(\recf,\vec{x}),$$ where $\Phi:\rr^\dd\times\dd\to\rr$ is
used to define $\recf(\vec{x})$ in terms of other values of $\recf$.
In this paper we consider functions $\recf$ that take natural numbers
as arguments but output real values, e.g., corresponding to costs such
as energy consumption, which need not be measured as integer values in
practice. Working with real values also makes optimizing for the
regression coefficients easier.
We restrict ourselves to the domain $\naturalnumbers^{\numparam}$
because in our motivating application, cost/size analysis, the input
arguments to recurrences represent data sizes, which take non-negative
integer values.
Technically, our approach may be easily extended to recurrence
equations on functions with domain $\integernumbers^{\numparam}$.

A system of recurrence equations is a functional equation on multiple
$f_i:\dd_i\to\rr$,
i.e., $\forall i,\,\forall {\vec{x}\in\dd_i},\; f_i(\vec{x}) = \Phi_i(f_1,\ldots\!,f_r,\vec{x})$.
Inequations and non-deterministic equations can also be considered by
making $\Phi$ non-deterministic, i.e.,
$\Phi:\rr^\dd\times\dd\to\pp(\rr)$ and $\forall
{\vec{x}\in\dd},\;f(\vec{x}) \in \Phi(f,\vec{x})$.  In general, such
equations may have any number of solutions.
In this work, we focus on deterministic recurrence equations, as we
will discuss later.

Classical recurrence relations of order $k$ are recurrence relations where
$\dd=\nn$, $\Phi(f,n)$ is a constant when $n < k$ and $\Phi(f,n)$ depends only
on $f(n-k),\ldots\!,f(n-1)$ when $n \geq k$.
For example, the following recurrence relation of order $k=1$, where
$\Phi(f,n)= f(n-1) + 1$ when $n \geq 1$, and $\Phi(f,n)= 1$ when $n <
1$ (or equivalently, $n = 0$), i.e.,
\begin{equation}
\label{eq:fibo-recurrence}
f(n) =
  \begin{cases*}
    1 & if $n = 0$, \\
    f(n-1) +  1 &  if $n \geq 1$,
  \end{cases*}
\end{equation}

\noindent
has the closed-form function $f(n) = n + 1$ as a solution.

Another example, with order $k=2$, is the historical Fibonacci
recurrence relation, where $\Phi(f,n)=f(n-1)+f(n-2)$ when $n\geq 2$,
and $\Phi(f,n)= 1$ when $n < 2$:
\begin{equation}
\label{eq:fibo-recurrence}
f(n) =
  \begin{cases*}
    1 & if $n = 0$ or $n = 1$, \\
    f(n-1) + f(n-2) &  if $n \geq 2$. \\
  \end{cases*}
\end{equation}

$\Phi$ may be viewed as a \emph{recursive definition} of ``the'' solution $f$ of
the equation, with the caveat that the definition may be non-satisfiable or
partial (degrees of freedom may remain). We define below an \emph{evaluation
strategy} $\mathsf{EvalFun}$ of this recursive definition, which, when it
terminates for all inputs, provides \emph{a} solution to the equation. This will
allow us to view recurrence equations as programs that can be evaluated,
enabling us to easily generate input-output pairs, on which we can perform
regression to attempt to guess a symbolic solution to the equation.

This setting can be generalized to \emph{partial solutions} to the equation,
i.e., partial functions $f:\dd\rightharpoonup\rr$ such that $f(\vec{x}) =
\Phi(f,\vec{x})$ whenever they are defined at $\vec{x}$.

We are interested in \emph{constrained recurrence relations}, where $\Phi$ is
expressed piecewise as
\begin{equation}
  \label{eq:new1}
  \Phi(f,\vec{x}) =
  \begin{cases*}
    e_1(\vec{x}) & if $\varphi_1(\vec{x})$ \\
    e_2(\vec{x}) & if $\varphi_2(\vec{x})$ \\
    \phantom{e_i(}\vdots & \phantom{$\varphi_i($}\vdots \\
    e_k(\vec{x}) & if $\varphi_k(\vec{x})$, \\
  \end{cases*}
\end{equation}
\noindent
with $f:\dd\to\rr$,
$\dd = \{\vec{x}\,|\,\vec{x} \in \nn^k \wedge \varphi_{\text{pre}}(\vec{x})\}$ for
some boolean constraint $\varphi_{\text{pre}}$, called the \emph{precondition of
$f$}, and $e_i(\vec{x})$, $\varphi_i(\vec{x})$ are respectively arbitrary
expressions and constraints over both $\vec{x}$ and $f$.
We further require that $\Phi$ is always defined, i.e.,
$\varphi_{\text{pre}} \models \vee_i\,\varphi_i$.
A case such that $e_i,\,\varphi_i$ do not contain any call to $f$ is called a
\emph{base case}, and those that do are called \emph{recursive cases}. In
practice, we are only interested in equations with at least one base case and
one recursive case.

A challenging class of recurrences that can be tackled with our approach is that
of ``nested'' recurrences, where recursive cases may contain nested calls
$f(f(...))$.

We assume that the $\varphi_i$ are mutually exclusive, so that $\Phi$
must be deterministic. This is not an important limitation for our
motivating application, cost analysis, and in particular the one
performed by the CiaoPP system. Such cost analysis can deal with a large class of
non-deterministic programs by translating the resulting
non-deterministic recurrences into deterministic ones. For example,
assume that the cost analysis generates the following recurrence
(which represents an input/output size relation).
\begin{equation*}
\begin{array}{ll}
 f(x) \, = \; 0 &  \text{ if } x = 0 \\
 f(x) \, = \; f(x-1) + 1 &  \text{ if } x > 0 \\
 f(x) \, = \; f(x-1) + 2 &  \text{ if } x > 0
\end{array}
\end{equation*}

\noindent
Then, prior to calling the solver, the recurrence is transformed into
the following two deterministic recurrences, the solution of which
would be an upper or lower bound on the solution to the original
recurrence. For upper bounds:
\begin{equation*}
\begin{array}{ll}
 f(x) \, = \; 0 &  \text{ if } x = 0, \\
 f(x) \, = \; \max(f(x-1) + 1, f(x-1) + 2) &  \text{ if } x > 0,
\end{array}
\end{equation*}
\noindent
and for lower bounds:
\begin{equation*}
\begin{array}{ll}
 f(x) \, = \; 0 &  \text{ if } x = 0, \\
 f(x) \, = \; \min(f(x-1) + 1, f(x-1) + 2) &  \text{ if } x > 0.
\end{array}
\end{equation*}

\noindent
Our regression technique correctly infers the solution $f(x) = 2 x$
in the first case, and $f(x) = x$ in the second case.
We have access to such program transformation, that recovers
determinism by looking for worst/best cases, under some hypotheses,
outside of the scope of this paper, on the kind of non-determinism and
equations that are being dealt with.

We now introduce an evaluation strategy of recurrences that allows us
to be more consistent with the termination of programs than the more
classical semantics consisting only of maximally defined partial
solutions. Let $\mathsf{def}(\Phi)$ denote a sequence $\langle
(e_1(\vec{x}),\varphi_1(\vec{x})),\ldots,(e_k(\vec{x}),\varphi_k(\vec{x}))\rangle$
defining a (piecewise) constrained recurrence relation $\Phi$ on $f$,
where each element of the sequence is a pair representing a case.  The
evaluation of the equation for a concrete value $\vvec{d}$, denoted
$\mathsf{EvalFun}(f(\vvec{d}))$, is defined as follows.
\begin{equation*}
  \mathsf{EvalFun}\big(f(\vvec{d})\,\big) = \mathsf{EvalBody}(\mathsf{def}(\Phi), \vvec{d}\,) \\
\end{equation*}
\begin{equation*}
  \mathsf{EvalBody}\big(\langle (e,\varphi)|\mathsf{Ps}\rangle, \vvec{d}\,\big) =
  \begin{cases*}
    \eval{e}{\vvec{d}}
    & if $\varphi\big(\vvec{d}\,\big)$ \\
    \mathsf{EvalBody}(\mathsf{Ps}, \vvec{d}) & if $\neg \varphi\big(\vvec{d}\,\big)$ \\
  \end{cases*}
\end{equation*}

The goal of our regression strategy is to find an expression $\candf$
representing a function $\dd\to\rr$ such that, for all $\vvec{d} \in \dd$,
\begin{itemize}
\item $\text{ If }\mathsf{EvalFun}(f(\vvec{d})) \text{ terminates, then }
  \mathsf{EvalFun}(f(\vvec{d})) =
\eval{\candf}{\vvec{d}}$,
  and
\item $\candf$ does not contain any recursive call in its definition.
\end{itemize}

If the above conditions are met, we say that $\candf$ is a \emph{closed form}
for $f$.
In the case of (sparse) linear regression, we are looking for
expressions
\begin{equation}
  \label{eq:2}
  \candf(\vec{x}) = \beta_0 + \beta_1 \ t_1(\vec{x}) + \beta_2 \ t_2(\vec{x}) + \ldots +
  \beta_{\numfeatures} \ t_{\numfeatures}(\vec{x}),
\end{equation}
where $\beta_i \in \rr$, and $t_i$ are expressions over $\vec{x}$, not including
recursive references to $\candf$.

For example, consider the following Prolog program which does not
terminate for a call \texttt{q(X)} where \texttt{X} is bound to a
non-zero integer.

\begin{minipage}[c]{0.95\textwidth}
\prettylstciao
  \begin{lstlisting}
q(X) :- X > 0, X1 is X + 1, q(X1).
q(X) :- X = 0.
  \end{lstlisting}
\end{minipage}

\noindent
The following
recurrence relation for its cost (in resolution steps) can be set up.
\begin{equation}
  \label{fig:cost-no-terminate}
  \begin{array}{ll}
    \stdub{\mathtt{q}}{x} \, = \; 1 &  \text{ if } x = 0 \\ [-1mm]
    \stdub{\mathtt{q}}{x} \, = \; 1 + \stdub{\mathtt{q}}{x+1} &  \text{ if } x > 0 \\ [-1mm]
  \end{array}
\end{equation}
\noindent
A CAS will give the closed form $\stdub{\mathtt{q}}{x} = 1 - x$ for
such recurrence, however, the cost analysis should give
$\stdub{\mathtt{q}}{x} = \infty$ for $x>0$.

\paragraph[(Sparse) Linear Regression]{\textbf{(Sparse) Linear Regression}.}
Linear regression~\citep{Hastie-09a} is a statistical technique used to
approximate the linear relationship between a number of explanatory (input)
variables and a dependent (output) variable. Given a vector of
(input) real-valued variables $X = (X_1,\ldots,X_{\numfeatures})^T$,
we predict the output variable $Y$ via the model
\begin{equation}
  \label{eq:2-2} \hat{Y} = \beta_0 + \sum_{i=1}^{\numfeatures}\beta_i X_i,
\end{equation}
which is defined through the vector of coefficients $\vec{\beta} =
(\beta_0,\ldots,\beta_{\numfeatures})^T \in \realnumbers^{\numfeatures+1}$.
Such coefficients are estimated from a set of observations
$\{(\langle x_{i1},\ldots,x_{i\numfeatures}\rangle,y_i)\}_{i=1}^{\numtraining}$ so as to minimize a loss function,
most commonly the sum of squares
\begin{equation}
  \label{eq:4} \vec{\beta} =
\underset{\vec{\beta} \in \realnumbers^{\numfeatures+1}}{\argmin}\sum_{i=1}^{\numtraining}\bigg(y_i - \beta_0 -
\sum_{j=1}^{\numfeatures} x_{ij}\beta_j\bigg)^2.
\end{equation}
Sometimes (as is our case) some of the input variables are not relevant to explain the output, but the above least-squares estimate will almost always assign nonzero values to all the coefficients. In order to force the estimate to make exactly zero the coefficients of irrelevant variables (hence removing them and doing \emph{feature selection}), various techniques have been proposed. The most widely used one is the lasso~\citep{Hastie-15a}, which adds an $\ell_1$ penalty on $\vec{\beta}$ (i.e., the sum of absolute values of each coefficient) to
Expression~\ref{eq:4}, obtaining
\vspace{-0.5em} 
\begin{equation}
  \label{eq:4-1} \vec{\beta} =
\underset{\vec{\beta} \in \realnumbers^{\numfeatures+1}}{\argmin}\sum_{i=1}^{\numtraining}\bigg(y_i - \beta_0 -
\sum_{j=1}^{\numfeatures} x_{ij}\beta_j\bigg)^2 + \lambda\,\sum_{j=1}^{\numfeatures}|\beta_j|,
\end{equation}
where $\lambda \ge 0$ is a hyperparameter that determines the level
of penalization: the greater $\lambda$, the greater the number of
coefficients that are exactly equal to $0$.
The lasso has two advantages over other feature selection techniques
for linear regression. First, it defines a convex problem whose unique
solution can be efficiently computed even for datasets where either of
$\numtraining$ or $\numfeatures$ are large (almost as efficiently as a standard linear
regression). Second, it has been shown in practice to be very good at
estimating the relevant variables.
In fact, the regression problem we would really like to solve is that of
Expression~\ref{eq:4} but subject to the constraint that
$\norm{(\beta_1,\ldots,\beta_{\numfeatures})^T}_0 \le K$, i.e., that at most $K$ of the $\numfeatures$
coefficients are non-zero, an $\ell_0$-constrained problem. Unfortunately,
this is an NP-hard problem. However, replacing the $\ell_0$ pseudo-norm
with the $\ell_1$-norm has been observed to produce good approximations in
practice~\citep{Hastie-15a}.

\paragraph[Symbolic Regression]{\textbf{Symbolic Regression}.}
Symbolic regression %
is a regression task in which the model space consists of all mathematical
expressions on a chosen signature, i.e., expression trees with variables or
constants for leaves and operators for internal nodes.
To avoid overfitting, objective functions are designed to penalize model
complexity, in a similar fashion to sparse linear regression techniques.
This task is much more ambitious: rather than searching over the vector space
spanned by a relatively small set of base functions as we do in sparse linear
regression, the search space is enormous, considering any possible expression
which results from applying a set of mathematical operators in any combination.
For this reason, heuristics such as evolutionary algorithms are typically used
to search this space, but runtime still remains a challenge for deep expressions.

The approach presented in this paper is parametric in the regression technique
used, and we instantiate it with both (sparse) linear and symbolic regression in our
experiments (Section~\ref{sec:implem-and-eval}). We will see that symbolic
regression addresses some of the limitations of (sparse) linear regression, at the
expense of time.
Our implementation is based on the symbolic regression library
\texttt{PySR}~\citep{pysr}, a multi-population evolutionary algorithm paired with
classical optimization algorithms to select constants.
In order to avoid overfitting and guide the search, \texttt{PySR}
penalizes model complexity, defined as a sum of individual node costs for
nodes in the expression tree, where a predefined cost is assigned to each
possible type of node.

In our application context, (sparse) linear regression searches for a solution to the
recurrence equation in the affine space spanned by candidate functions, i.e.,
${\candf = \beta_0 + \sum_i \beta_i t_i}$ with ${t_i\in \basefuns}$, while symbolic
regression may choose any expression built on the chosen operators.
For example, consider equation \texttt{exp3} of Table~\ref{table:newexp2}, whose
solution is $(x,y)\mapsto 2^{x+y}$. This solution cannot be expressed using
(sparse) linear regression and the set of candidates $\{\lambda xy.x,\lambda xy.y,\lambda
xy.x^2,\lambda xy.y^2,\lambda xy.2^x,\lambda xy.2^y\}$, but can be found with
symbolic regression and operators $\{+(\cdot,\cdot), \times(\cdot,\cdot),
2^{(\cdot)}\}$.

\section{Algorithmic Description of the Approach}
\label{sec:description-approach}

In this section we describe our approach for generating and checking
candidate solutions for recurrences that arise in resource analysis.

\subsection{A first version}
\label{sec:description-approach-v1}
Algorithms~\ref{algo1} and~\ref{algo2} correspond to the
\emph{guesser} and \emph{checker} components, respectively, which are
illustrated in Figure~\ref{fig:mlresol-arch}.
For the sake of presentation, Algorithm~\ref{algo1} describes the
instantiation of the guess method based on lasso linear regression.
It receives a recurrence relation for a function
$\recf$ to solve, a set of
base functions,
and a threshold to decide
when to discard irrelevant terms. The output is a closed-form
expression $\candf$ for $\recf$, and a \emph{score} $\score$ that reflects the
accuracy of the approximation, in the range $[0,1]$. If $\score
\approx 1$, the
approximation can be considered a candidate solution. Otherwise,
$\candf$ is
an approximation (not necessarily a bound) of the solution.
\paragraph[Generate]{\textbf{Generate}.} In line~\ref{algo1:l1} we start by
generating a set $\inputset$ of random inputs for $\recf$. Each input $\vec{x_i}$ is an
$\numparam$-tuple verifying precondition $\varphi_{\text{pre}}$, where $\numparam$ is the
number of arguments of $\recf$. In line~\ref{algo1:l2} we produce the
training set $\trainingset$.  The (explanatory) inputs are generated by evaluating the
base functions
in $\basefuns = \tuple{t_1, t_2, \ldots, t_{\numfeatures}}$ with each
tuple $\vec{x} \in \inputset$. This is done by using function $\evalfun$,
defined as follows.
\begin{equation*}
\evalfun(\tuple{t_1, t_2, \ldots, t_{\numfeatures}}, \vec{x}) =
\tuple{t_1(\vec{x}), t_2(\vec{x}), \ldots, t_{\numfeatures}(\vec{x})}
\end{equation*}
\noindent
We also evaluate the recurrence equation for input $\vec{x}$,
and add the observed output $\recf(\vec{x})$ as the
last element in the
corresponding training case.
\paragraph[Regress]{\textbf{Regress}.} In line~\ref{algo1:l3} we generate a
first linear model by applying function $\mathsf{CVLassoRegression}$ to the
generated training set,
which performs a sparse linear
regression with lasso regularization. As already mentioned, lasso
regularization requires a hyperparameter $\lambda$ that determines the
level of penalization for the coefficients. Instead of using a single
value for $\lambda$, $\mathsf{CVLassoRegression}$ uses a range of possible
values, applying cross-validation on top of the linear regression to
automatically select the best value for that parameter, from the given
range.
In particular, $k$-fold cross-validation is performed,
which means that the training set is split into $k$ parts or
\emph{folds}. Then, each fold is taken as the validation set, training
the model with the remaining $k-1$ folds. Finally, the performance
measure reported is the average of the values computed in the $k$
iterations.
The result of this function is the vector of coefficients
$\vec{\beta^{\prime}}$, together with
the intercept
$\beta_0^{\prime}$. These coefficients are used in line~\ref{algo1:l4}
to decide which base functions
are discarded before the last regression
step. Note that $\mathsf{RemoveTerms}$
removes the base functions
from $\basefuns$
together with their corresponding output values from the training set
$\trainingset$, returning the new set of base functions
$\basefuns^{\prime}$
and its corresponding training set $\trainingset^{\prime}$. In
line~\ref{algo1:l5}, standard linear regression (without
regularization or cross-validation) is applied, obtaining the final
coefficients $\vec{\beta}$ and $\beta_0$. Additionally, from this step
we also obtain the score $\score$ of the resulting model. In
line~\ref{algo1:l6} we set up the resulting closed-form expression,
given as a function on the variables in $\vec{x}$. Note that we use
the function $E$ to bind the variables in the base functions
to the arguments of the closed-form expression. Finally, the
closed-form expression and its corresponding score are returned as the
result of the algorithm.

\begin{figure*}
  \centering
  \begin{algorithm}[H]
  \label{algo1}
    \SetKwInOut{Input}{Input}
    \SetKwInOut{Output}{Output}

    \Input{Target recurrence relation on $\recf: \dd \to \realnumbers$.
      \newline
      $\varphi_{\text{pre}}$: precondition defining $\dd$.
      \newline
      $\numtraining$: number of (random) inputs for evaluating $\recf$.
      \newline
      $\basefuns \subseteq \dd \to \realnumbers$: set of base functions (represented as a tuple).
      \newline
      $\Lambda$: range of values to automatically choose a
      lasso hyperparameter $\lambda \in \realnumbers^+$ that
      maximizes the performance of the model via cross-validation.
      \newline
      $k$: indicates performing $k$-fold cross-validation, $k \geq 2$.
      \newline
      $\epsilon \in \realnumbers^+$: threshold for term ($t_i \in \basefuns$) selection.
    }
    \Output{$\candf \in \mathsf{Exp}$: a candidate solution (or an approximation) for $\recf$.
      \newline
      $\score \in [0,1]$: score, indicating the accuracy of the estimation ($R^2$).
    }

    $\inputset \gets \{\vec{x_i}\,|\,\vec{x_i} \in \nn^{\numparam} \wedge \varphi_{\text{pre}}(\vec{x_i}) \}_{i=1}^\numtraining$ \tcp*{$\numtraining$ random inputs for $f$} \label{algo1:l1}
    $\trainingset \gets \{(E(\basefuns,\vec{x}),\recf(\vec{x})) \,|\,
    \vec{x} \in \inputset \}$ \tcp*{Training set} \label{algo1:l2}
    $( \vec{\beta^{\prime}},\beta^{\prime}_0) \gets
    \mathsf{CVLassoRegression}(\trainingset, \Lambda, k)$\;  \label{algo1:l3}
    $(\basefuns^{\prime},\trainingset^{\prime}) \gets
    \mathsf{RemoveTerms}(\basefuns,\trainingset,\vec{\beta^{\prime}},\beta_0^{\prime},\epsilon)$
    \tcp*{$\epsilon$-pruning} \label{algo1:l4}
      $(\vec{\beta},\beta_0,\score) \gets
      \mathsf{LinearRegression}(\trainingset^{\prime})$\; \label{algo1:l5}
    $\candf \gets \beta_0 + \lambda \vec{x} \cdot \vec{\beta}^{T} \times
    E(\basefuns^{\prime},\vec{x})$\; \label{algo1:l6}
    \Return{$(\candf, \score)$}\;
    \caption{Candidate Solution Generation (\emph{Guesser}).}
  \end{algorithm}
\end{figure*}

\paragraph[Verify]{\textbf{Verify}.}

\begin{figure*}[t]
  \centering
  \begin{algorithm}[H]
  \label{algo2}
    \SetKwInOut{Input}{Input}
    \SetKwInOut{Output}{Output}

    \Input{Target recurrence relation on $\recf:\dd \to \realnumbers$.
      \newline
      $\varphi_{\text{pre}}$: precondition defining $\dd$.
      \newline
      $\candf \in \mathsf{Exp}$: a candidate solution for $\recf$.
    }
    \Output{$\mathsf{true}$ if $\candf$ is a solution for $\recf$, $\mathsf{false}$ otherwise.
    }
    $\varphi_{\text{previous}} \gets \mathsf{true}$ \; $\mathsf{Formula} \gets \mathsf{true}$ \;
    \ForEach{$(e,\varphi) \in \mathsf{def}(\recf)$}{
      $\mathsf{Eq} \gets \mathsf{replaceCalls}(``\recf(\vec{x}) - e = 0",\recf(\vec{x}),\candf,\varphi_{\text{pre}},\varphi)$\; \label{linereplacecalls}
      \eIf{$\neg \: \mathsf{containsCalls}(\mathsf{Eq},\recf)$}{
      $\mathsf{Eq} \gets \mathsf{simplifyCAS}(\mathsf{inlineCalls}(\mathsf{Eq},\candf,\mathsf{def}(\candf)))$\; \label{linesimpl}
      \eIf{$\mathsf{supportedSMT}(\mathsf{Eq})$}{
        $\mathsf{Formula} \gets ``\mathsf{Formula} \wedge (\varphi_{\text{pre}} \wedge \varphi_{\text{previous}} \wedge \varphi \implies \mathsf{Eq})"$\;
        $\varphi_{\text{previous}} \gets ``\varphi_{\text{previous}} \wedge \neg \varphi"$ \;
      }{
        \Return{$\mathsf{false}$}\;
      }}{
      \Return{$\mathsf{false}$}\;
    }

      }

    \Return{$(\not\models_{\text{SMT}} \llbracket \neg \mathsf{Formula} \rrbracket_{\text{SMT}})$}\; \label{lastline}
    \caption{Solution Checking (\emph{Checker}).}
  \end{algorithm}
\end{figure*}

Algorithm~\ref{algo2}
mainly relies on an SMT-solver and a CAS.
Concretely, given the
constrained recurrence relation on $\recf:\dd \to \rr$ defined by
\begin{equation*}
  \label{eq:1-1}
  \recf(\vec{x}) = \small
  \begin{cases*}
    e_1(\vec{x}) & if $\varphi_1(\vec{x})$ \\
    e_2(\vec{x}) & if $\varphi_2(\vec{x})$ \\
    \phantom{e_k}\vdots  & \phantom{if $\varphi\!$}\vdots \\
    e_k(\vec{x}) & if $\varphi_k(\vec{x})$ \\
  \end{cases*}
\end{equation*}
our algorithm constructs the logic formula
\begin{equation}
  \label{smtrep}
  \bigg \llbracket \bigwedge\limits_{i=1}^k \left(\left(\bigwedge\limits_{j=1}^{i-1} \neg  \varphi_j(\vec{x}) \right)  \wedge \varphi_i(\vec{x}) \wedge \varphi_{\text{pre}}(\vec{x}) \implies \mathsf{Eq}_i \right) \bigg \rrbracket_{\text{SMT}}
\end{equation}
where operation $\llbracket e \rrbracket_{\text{SMT}}$ is the
translation of any expression $e$ to an SMT-LIB expression, and
$\mathsf{Eq}_i$ is the result of replacing in $\recf(\vec{x}) =
e_i(\vec{x})$ each occurrence of $\recf$ (\emph{a priori} written as
an uninterpreted function) by the definition of the candidate
solution $\candf$ (by using $\mathsf{replaceCalls}$ in
line~\ref{linereplacecalls}), and performing a simplification
(by using $\mathsf{simplifyCAS}$ in line~\ref{linesimpl}).
The function
$\mathsf{replaceCalls}(\mathsf{expr},\recf(\vec{x}^\prime),\candf,\varphi_{\text{pre}},\varphi)$
replaces every subexpression in $\mathsf{expr}$ of the form
$\recf(\vec{x}^\prime)$ by $\candf(\vec{x}^\prime)$, if
$\varphi_{\text{pre}}(\vec{x}^\prime) \wedge \varphi \implies
\varphi_{\text{pre}}(\vec{x}^\prime)$.
When checking Formula~\ref{smtrep}, all variables are assumed to be
integers. As an implementation detail, to work with \texttt{Z3} and
access a large arithmetic language (including rational division),
variables are initially declared as reals, but integrality constraints
$\large(\bigwedge_i \vec{x}_i = \lfloor \vec{x}_i \rfloor\large)$
are added to the final formula.
Note that this encoding is consistent with the evaluation
($\mathsf{EvalFun}$) described in Section~\ref{sec:prelim}.

The goal of $\mathsf{simplifyCAS}$ is to obtain (sub)expressions
supported by the SMT-solver. This typically allows simplifying away
differences of transcendental functions, such as exponentials and
logarithms, for which SMT-solvers like Z3 currently have extremely
limited support, often dealing with them as if they were uninterpreted
functions.  For example, \texttt{log2} is simply absent from SMT-LIB,
although it can be modelled with exponentials, and reasoning abilities
with exponentials are very limited: while Z3 can check that $2^x - 2^x
= 0$, it cannot check (without further help) that
$2^{x+1}-2\cdot2^x=0$.  Using $\mathsf{simplifyCAS}$, the latter is
replaced by $0=0$ which is immediately checked.

Finally, the algorithm asks the SMT-solver
for models of the negated formula (line~\ref{lastline}). If no model
exists, then it returns $\mathsf{true}$, concluding that $\candf$ is an exact
solution to the recurrence, i.e., $\candf(\vec{x}) = \recf(\vec{x})$ for
any input $\vec{x} \in \dd$ such that $\mathsf{EvalFun}(\recf(\vec{x}))$
terminates. Otherwise, it returns $\mathsf{false}$.
Note that, if we are not able to express $\candf$ using the syntax
supported by the SMT-solver, even after performing the simplification
by $\mathsf{simplifyCAS}$, then the algorithm finishes returning
$\mathsf{false}$.

\subsection{Extension: domain splitting}
\label{sec:description-approach-domain-splitting}

For the sake of exposition, we have only presented a basic combination of
Algorithms~\ref{algo1} and \ref{algo2}, but the core approach \emph{generate,
regress and verify} can be generalized to obtain more
accurate results.
Beyond the use of different regression algorithms (replacing
\mbox{lines~\ref{algo1:l3}--\ref{algo1:l6}} in Algorithm~\ref{algo1}, e.g., with
symbolic regression as presented in Section~\ref{sec:prelim}), we can also
decide to apply Algorithm~\ref{algo1} separately on multiple subdomains of
$\dd$: we call this strategy \emph{domain splitting}.
In other words, rather than trying to directly infer a solution on $\dd$
by regression, we do the following.
\begin{itemize}
  \item \emph{Partition} $\dd$ into subdomains $\dd_i$.
  \item Apply (any variant of) Algorithm~\ref{algo1} on each $\dd_i$,
    i.e., \emph{generate} input-output pairs, and \emph{regress} to obtain
    candidates $\candf_i:\dd_i\to\realnumbers$.
  \item This gives a global candidate
    $\candf : x \mapsto \{\candf_i \text{ if } x\in\dd_i\}$,
    that we can then attempt to \emph{verify} (Algorithm~\ref{algo2}).
\end{itemize}

A motivation for doing so is the observation that it is easier for regression
algorithms to discover expressions of ``low (model) complexity'', i.e., expressions of low
depth on common operators for symbolic regression and affine combinations of
common functions for (sparse) linear regression
(note that \emph{model complexity of the expression} is not to be confused with
the \emph{computational complexity of the algorithm} whose cost is
represented by the expression, i.e., the asymptotic rate of growth of
the corresponding function).
We also observe that our equations often admit solutions that can be described
piecewise with low (model) complexity expressions, where equivalent global expressions
are somehow artificial: they have high (model) complexity, making them hard or
impossible to find.
In other words, equations with ``piecewise simple'' solutions are more common
than those with ``globally simple'' solutions, and the domain splitting strategy
is able to decrease complexity for this common case by reduction to the more
favorable one.

For example, consider equation \texttt{merge} in Table~\ref{table:newexp2},
representing the cost of a merge function in a merge-sort algorithm. Its
solution is
\begin{align*}
  f : \naturalnumbers^2 &\to \realnumbers\\
      (x,y) &\mapsto
  \begin{cases*}
    x+y-1 & if $x>0 \land y>0$, \\
    0     & if $x=0 \lor  y=0$.
  \end{cases*}
\end{align*}
It admits a piecewise affine description, but no simple global description
as a polynomial, although we can admittedly write it as
$\min(x,1)\times\min(y,1)\times(x+y-1)$, which is unreachable by (sparse) linear
regression for any reasonable set of candidates, and of challenging depth for
symbolic regression.
\\

To implement domain splitting, there are two challenges to face:
(1) partition the domain into appropriate subdomains that make regression more
accurate and
(2) generate random input points inside each subdomain.

In our implementation (Section~\ref{sec:implem-and-eval}), we test a
very simple version of this idea, were generation is handled by a trivial
rejection strategy, and where the domain is partitioned using the conditions
$\varphi_i(\vec{x})\wedge\bigwedge_{j=1}^{i-1}\neg\varphi_j(\vec{x})$
that define each clause of the input equation.

In other words, our splitting strategy is \emph{purely syntactic}. More advanced
strategies could learn better subdomains, e.g., by using a generalization of
model trees, and are left for future work. However, as we will see in
Section~\ref{sec:implem-and-eval}, our naive strategy already provides good
improvements compared to Section~\ref{sec:description-approach-v1}.
Intuitively, this seems to indicate that a large part of the
``disjunctive behavior'' of solutions originates from the ``input
disjunctivity'' in the equation (which, of course, can be found in other places
than the $\varphi_i$, but this is a reasonable first approximation).

Finally, for the verification step, we can simply construct an SMT formula
corresponding to the equation as in Section~\ref{sec:description-approach-v1},
using an expression defined by cases for $\candf$, e.g., with the \texttt{ite}
construction in SMT-LIB.

\section{Our Approach in the Context of Static Cost Analysis of (Logic) Programs}
\label{sec:application-cost-analysis}

In this section, we describe how our approach could be used in the
context of the motivating application, Static Cost Analysis.  Although
it is general, and could be integrated into any
cost analysis system based on recurrence solving, we illustrate
its use in the context of the \ciaopp system.
Using a logic program, we first illustrate how
\ciaopp
sets up recurrence relations representing the sizes of output
arguments of predicates and the cost of such predicates.  Then, we
show how our novel approach is used to solve a recurrence relation
that cannot be solved by \ciaopp.

\begin{example}
\label{ex:running-code-regr}
Consider predicate \verb-q/2- in Figure~\ref{fig:example1-code}, and
calls to it where the first argument is bound to a non-negative
integer and the second one is a free variable.\footnote{This set of
calls is represented by the ``entry'' assertion at the beginning of
the code, where property \texttt{nnegint} stands for the set of
non-negative integers.} Upon success of these calls, the second
argument is bound to an non-negative integer too. Such calling mode,
where the first argument is input and the second one is output, is
automatically inferred by \ciaopp
(see~\cite{ciaopp-sas03-journal-scp-short} and its references).
\end{example}

\begin{figure}[h]
  \centering
  \prettylstciao
  \begin{lstlisting}
:- entry q/2: nnegint*var.
q(X,0):-
  X=0.
q(X,Y):-
  X>0,
  X1 is X - 1,
  q(X1,Y1),
  q(Y1,Y2),
  Y is Y2 + 1.
\end{lstlisting}
  \caption{A program with a nested recursion.}
  \label{fig:example1-code}
\end{figure}

The \ciaopp{} system first infers size relations for the different
arguments of predicates, using a rich set of size metrics
(see~\cite{resource-iclp07-short,plai-resources-iclp14-short} for details). Assume
that the size metric used in this example, for the numeric argument
\texttt{X} is the \emph{actual value} of it (denoted \texttt{int(X)}).
The system will try to infer a function ${\tt S}_\mathtt{q}(x)$ that
gives the size of the output argument of \texttt{q/2} (the second
one), as a function of the size ($x$) of the input argument (the first
one). For this purpose, the following size relations for ${\tt
  S}_\mathtt{q}(x)$ are automatically set up (the same as the
recurrence in Equation~\ref{eq:recurrence} used in
Section~\ref{sec:overview} as example).
\begin{equation}
  \label{fig:size1}
  \begin{array}{ll}
    \sizeub{\mathtt{q}}{x} \, = \; 0 &  \text{ if } x = 0 \\ [-1mm]
    \sizeub{\mathtt{q}}{x} \, = \; \sizeub{\mathtt{q}}{\sizeub{\mathtt{q}}{x-1}}  + 1 &  \text{ if } x > 0 \\ [-1mm]
  \end{array}
\end{equation}
\noindent
The first and second recurrence correspond to the first and second
clauses respectively (i.e., base and recursive cases). Once recurrence
relations (either representing the size of terms, as the ones above,
or the computational cost of predicates, as the ones that we will see
later) have been set up, a solving process is started.

Nested recurrences, as the one that arise in this example, cannot be
handled by most state-of-the-art recurrence solvers. In particular,
the modular solver used by \ciaopp{} fails to find a closed-form
function for the recurrence relation above. In contrast, the novel
approach that we propose obtains the
closed form $\hat{{\tt S}}_\mathtt{q}(x) = x$, which is an
exact solution of such recurrence (as shown in Section~\ref{sec:overview}).

Once the size relations have been inferred, \ciaopp{} uses them
to infer the computational cost of a call to \texttt{q/2}. For
simplicity, assume that in this example, such cost is given in terms
of the number of \emph{resolution steps},
as a function of the size of the input argument, but note that
\ciaopp's cost analysis is parametric with respect to resources, which
can be defined by the user by means of a rich assertion language, so
that it can infer a wide range of resources, besides resolution steps.
Also for simplicity, we assume that
all builtin predicates, such as arithmetic/comparison operators
have zero cost (in practice there is a ``trust'' assertion for each
builtin that specifies its cost as if it had been inferred by the
analysis).

In order to infer the cost of a call to \texttt{q/2}, represented as
$\stdub{\mathtt{q}}{x}$, \ciaopp{} sets up the following
\costrelations{}, by using the size relations inferred previously.
\begin{equation}
  \label{fig:cost1}
  \begin{array}{ll}
    \stdub{\mathtt{q}}{x}\, = \; 1 &  \text{ if } x = 0 \\ [-1mm]
    \stdub{\mathtt{q}}{x}\, = \; \stdub{\mathtt{q}}{x-1} + \stdub{\mathtt{q}}{\sizeub{\mathtt{q}}{x-1}}  + 1 &  \text{ if } x > 0 \\ [-1mm]
  \end{array}
\end{equation}
\noindent
We can see that the cost
of the second recursive call to predicate \texttt{p/2} depends on the
size of the output argument of the first recursive call to such
predicate, which is given by function $\sizeub{\mathtt{q}}{x}$, whose
closed form $\sizeub{\mathtt{q}}{x} = x$ is computed by our approach,
as already explained.
Plugging such closed form into the
recurrence relation above, it can now be solved by \ciaopp, obtaining
$\stdub{\mathtt{q}}{x} = 2^{x+1} - 1$.

\section{Implementation and Experimental Evaluation}
\label{sec:implem-and-eval}

We have implemented a prototype of our novel approach and performed an
experimental evaluation in the context of the \ciaopp system, by
solving recurrences similar to those generated during static cost
analysis, and comparing the results with the current \ciaopp solver as
well as with state-of-the-art cost analyzers and recurrence
solvers. Our experimental results are summarized in
Table~\ref{table:newexp1} and Figure~\ref{fig:barchart}, where our
approach is evaluated on the benchmarks of Table~\ref{table:newexp2},
and where we compare its results with other tools, as described below
and in Section~\ref{sec:related-work}. Beyond these summaries, additional
details on chosen benchmarks are given in paragraph~\emph{Evaluation
and Discussion}, and full outputs are included
in~\ref{appendix:full-output}.

\paragraph[Prototype]{\textbf{Prototype}.}

As mentioned earlier, our approach is parameterized by the regression
technique employed, and we have instantiated it with both (sparse) linear and
symbolic regression. To assess the performance of each individually,
as well as their combinations with other techniques, we have developed
three versions of our prototype, which
appear as the first three columns (resp. bars) of
Table~\ref{table:newexp1} (resp. Figure~\ref{fig:barchart}, starting
from the top).  The
simplest version of our approach, as described in
Section~\ref{sec:description-approach-v1}, is \texttt{mlsolve(L)}.  It
uses linear regression with lasso regularization and feature
selection.  \texttt{mlsolve(L)+domsplit} is the same tool, adding the
domain splitting strategy described in
Section~\ref{sec:description-approach-domain-splitting}.  Finally,
\texttt{mlsolve(S)+domsplit} replaces (sparse) linear regression with symbolic
regression.

The overall
architecture, input and outputs of our prototypes were already
illustrated in Figure~\ref{fig:mlresol-arch} and described in
Sections~\ref{sec:overview} and~\ref{sec:description-approach}.
The implementation is written in \texttt{Python 3.6/3.7}, using
\texttt{Sympy}~\citep{sympy} as CAS,
\texttt{Scikit-Learn}~\citep{scikit-learn} for regularized linear
regression and \texttt{PySR}~\citep{pysr} for symbolic regression.
We use \texttt{Z3}~\citep{z3} as SMT-solver, and
\texttt{Z3Py}~\citep{z3py} as interface.

\paragraph[Legend]{\textbf{Legend}.}
Columns $3$--$19$ of Table~\ref{table:newexp1}
are categorized, from left to right, as 1) the prototypes of this
paper's approach, 2) \ciaopp builtin solver, 3) other program analysis
tools, and 4) Computer Algebra Systems (CAS) with recurrence
solvers.
Such solvers also appear in the $y$-axis of Figure~\ref{fig:barchart},
and there is a horizontal bar (among \texttt{a}-\texttt{q}) for each
of them.

From top to bottom, our benchmarks are organized by category
reflecting some of their main features. Benchmarks are manually
translated into the input language of each tool, following a
best-effort approach. The symbol ``\texttt{(*)}'' corresponds to help
provided for certain imperative-oriented tools.
More comments will be given in
\emph{tools} and \emph{benchmarks} paragraphs below.

For the sake of simplicity, and for space reasons, we report the
accuracy of the results by categorizing them using only $4$
symbols. ``\good'' corresponds to an exact solution, ``\approxbnd'' to
a ``precise approximation'' of the solution, ``\bad'' to another
non-trivial output, and ``\none'' is used in all remaining cases
(trivial
infinite bound, no
output, error, timeout, unsupported). Nevertheless, the concrete
functions output are included in~\ref{appendix:full-output}.
Figure~\ref{fig:barchart} also shows the number of closed-form
functions inferred broken down by these four accuracy categories, in
the form of stacked horizontal bars.
In addition, when a solution $f$ has superpolynomial growth, we relax
our requirements and may write ``$e^\theta$'' whenever $\candf$ is
not a precise approximation of $f$, but $\log{\candf}$ is a precise
approximation of $\log{f}$.\footnote{This accounts for the fact that
approximation classes of functions with superpolynomial growth are too
thin for our purposes: $x^{1.619}$ is not a precise approximation of
$x^{1.618}$.}
We include ``$e^\theta$'' in the ``\approxbnd'' category in
Figure~\ref{fig:barchart}.
Given the common clash of definitions when using $O$ notation on several
variables, we make explicit the notion of approximation used here.

We say that a function $\candf$ is a \emph{precise approximation} of
$f:\dd\to\realnumbers$ with $\dd\subset\naturalnumbers^\numparam$
whenever
$$\forall u\in\dd^\naturalnumbers,
  \Vert u_n \Vert\underset{n\to\infty}{\longrightarrow}\infty \implies
  \big(n\mapsto \candf(u_n)\big) \in \Theta\big(n\mapsto f(u_n)\big),$$
where $\Theta$ is the usual uncontroversial one-variable big theta,
and $\Vert\cdot\Vert$ is any norm on $\realnumbers^k$. %
In other words, a \emph{precise approximation} is a function which is
asymptotically within a constant factor of the approximated function,
where ``asymptotically'' means that some of the input values tend to
infinity, but not necessarily all of them.
Note that $\max(x,y)$ and $x+y$ belong to the same class, but $(x,y)\mapsto x+y-1$
is not a precise approximation of the solution of the \texttt{merge} benchmark
(set $u_n = (n,0)$): the approximations we consider give ``correct asymptotic
class in all directions''.

Hence, bounds that end up in the ``\bad''
category may still be precise enough
to be useful in the context of static analysis. We nevertheless resort to this
simplification to avoid the presentation of an inevitably complex lattice of
approximations.
It may also be noted that different tools do not judge quality in the same way.
CAS typically aim for \emph{exact} solutions only, while static analysis tools
often focus on provable \emph{bounds} (exact or asymptotic), which the notion of
approximation used in this table does not require. Once again, we use a single
measure for the sake of simplicity.

\begin{figure}[t]
  \vspace{-0.5em}
    \resizebox{0.97\textwidth}{!}{%
    \input{\figpath/precisionbarchart.pgf}
    }
  \vspace{-1.25em}
  \caption{Comparison of solver tools by accuracy of the result.}
  \label{fig:barchart}
  \raggedright
\end{figure}

\clearpage
\newpage

\renewcommand{\gape}[1]{\makecell[l]{\phantom{.}\\[-6.5pt]#1\\[-6.5pt]\phantom{.}}}

{\footnotesize
 \renewcommand{\arraystretch}{2}%
\begin{longtable}[c]{| c | c | p{5.75cm} | p{3.5cm} |}
\caption{Benchmarks\label{table:newexp2}}\\
\cline{1-4}
\textbf{Category} & \textbf{Bench} & \textbf{Equation} & \textbf{Solution} \\ \cline{1-4}
\endfirsthead
\caption[]{Benchmarks (continued)}\\
\cline{1-4}
\textbf{Category} & \textbf{Bench} & \textbf{Equation} & \textbf{Solution} \\ \cline{1-4}
\endhead
\multirow{6.5}{*}{\texttt{scale}}
 & \makecell[c]{
    \phantom{.}\\[-0.9em]
    \texttt{highdim1}}
 & \gape{\makecell[l]{
   \phantom{.}\\[-0.9em]
   \scalebox{0.75}{$\vec{x}=(x_1,\ldots,x_{10})$,}
   \\
   \scalebox{0.75}{$f(\vec{x})=
     \begin{cases*}
      i+f(\vec{x}[x_i\leftarrow (x_i-1)]) & if $x_i > 0 \land \forall j<i,\, x_j = 0$ \\[-2pt]
      0                                   & if $\forall i,\, x_i=0$
     \end{cases*}$}}}
 & $\sum_{k=1}^{10} kx_k$
 \\[-.75em] \cline{2-4}
 & \texttt{poly2}
 & $f_9(x,y)$? (cf. \texttt{highdim2})
 & $xy$
 \\ \cline{2-4}
 & \texttt{poly5}
 & $f_7(t,x,y,z)$? (cf. \texttt{highdim2})
 & $txyz$
 \\ \cline{2-4}
 & \texttt{poly7}
 & \gape{\scalebox{0.7}{\makecell[l]{
    $f_7(t,x,y,z)+f_9(t,x)+f_7(x,x,x,z)+f_{10}(y)+f_7(z,z,z,z)$?\\[.5em]
    (cf. \texttt{highdim2})
   }}}
 & $txyz+tx+x^3z+y+z^4$
 \\ \cline{2-4}
 & \texttt{highdim2}
 & \gape{\makecell[l]{
   \scalebox{0.7}{$\forall 1 \leq i \leq 10,\; f_i(x_i,\ldots,x_{10})=$}
   \\[.25em]
   \scalebox{0.7}{$\;\;\;
     \begin{cases*}
      f_{i+1}(x_{i+1},\ldots,x_{10})+f_i(x_i-1,x_{i+1},\ldots,x_{10}) & if $x_i > 0$ \\[-2pt]
      0                                                          & if $x_i = 0$, \\
     \end{cases*}$}
   \\[.5em]
   \scalebox{0.7}{$f_{11}() = 1$}
   }}
 & $f_1(\vec{x})=\prod_{k=1}^{10} x_k$
 \\ \cline{1-4}
\multirow{6}{*}{\texttt{amortized}}
 & \texttt{loop\_tarjan}
 &  \gape{\makecell[l]{
    \scalebox{.7}{Appropriate encoding of total number of loop iterations of}\\
    \;\;\lstinputlisting[language=c,basicstyle=\tiny\ttfamily,frameround=tttt,frame=single,
       xleftmargin=.2cm, xrightmargin=.2cm, linewidth=4cm]{loopus\_tarjan.c}\\
   }}
 & $2n$ (worst-case)
 \\ \cline{2-4}
 & \texttt{enqdeq1}
 & \gape{\scalebox{.7}{\makecell[l]{Appropriate encoding of total number of ticks of\\
   \;\;\texttt{queue = ([], []);}
   \texttt{Do $n$ enqueue;}
   \texttt{Do $n$ dequeue}\\
   (cf. Figure~\ref{fig:enqdeq})}}}
 & $3n$
 \\ \cline{2-4}
 & \texttt{enqdeq2}
 & \gape{\scalebox{.7}{\makecell[l]{Appropriate encoding of total number of ticks of\\
   \;\;\texttt{queue = ([], Lo);} (with \texttt{Lo} of length $k$)\\
   \;\;\texttt{Do $n$ enqueue;}
   \texttt{Do $m$ dequeue}\\
   (cf. Figure~\ref{fig:enqdeq})}}}
 & \gape{\scalebox{0.7}{$\begin{cases*}
      n+m  & if $m \leq k$ \\[-2pt]
      2n+m & if $m > k$   \\
    \end{cases*}$}}
 \\ \cline{2-4}
 & \texttt{enqdeq3}
 & \gape{\scalebox{.7}{\makecell[l]{Appropriate encoding of total number of ticks of\\
   \;\;\texttt{queue = ([], []);}
   \texttt{Do $2n$ enqueue;}\\
   \;\;\texttt{Do $n$ dequeue;}
   \texttt{Do $n$ enqueue;} \texttt{Do $n$ dequeue}\\
   (cf. Figure~\ref{fig:enqdeq})}}}
 & 7n
 \\ \cline{1-4}
\multirow{6.25}{*}{\texttt{max-heavy}}
 & \texttt{merge-sz}
 & \gape{\scalebox{0.7}{$f(x,y) =
    \begin{cases*}
      \max(f(x-1,y), f(x,y-1)) + 1 & if $x > 0 \land y > 0$ \\[-2pt]
      x                            & if $x > 0 \land y = 0$ \\[-2pt]
      y                            & if $x = 0 \land y > 0$ \\
    \end{cases*}$}}
 & $x+y$
 \\ \cline{2-4}
 & \texttt{merge}
 & \gape{\scalebox{0.7}{$f(x,y) =
    \begin{cases*}
      \max(f(x-1,y), f(x,y-1)) + 1 & if $x > 0 \land y > 0$ \\[-2pt]
      0                            & if $x = 0 \lor  y = 0$ \\
    \end{cases*}$}}
 & \gape{\scalebox{0.7}{$\begin{cases*}
      x+y-1 & if $x>0 \land y>0$ \\[-2pt]
      0     & if $x=0 \lor  y=0$ \\
    \end{cases*}$}}
 \\ \cline{2-4}
 & \texttt{open-zip}
 & \gape{\scalebox{0.7}{$f(x,y) =
    \begin{cases*}
      f(x-1,y-1) + 1  & if $x > 0 \land y > 0$ \\[-2pt]
      f(x,y-1)   + 1  & if $x = 0 \land y > 0$ \\[-2pt]
      f(x-1,y)   + 1  & if $x > 0 \land y = 0$ \\[-2pt]
      0               & if $x = 0 \land y = 0$ \\
    \end{cases*}$}}
 & $\max(x,y)$
 \\ \cline{2-4}
 & \texttt{s-max}
 & \gape{\scalebox{0.7}{$f(x,y) =
    \begin{cases*}
      \max(y,f(x-1,y)) + 1 & if $x > 0$ \\[-2pt]
      y                    & if $x = 0$ \\
    \end{cases*}$}}
 & $x+y$
 \\ \cline{2-4}
 & \texttt{s-max-1}
 & \gape{\scalebox{0.7}{$f(x,y) =
    \begin{cases*}
      \max(y,f(x-1,y+1)) + 1 & if $x > 0$ \\[-2pt]
      y                      & if $x = 0$ \\
    \end{cases*}$}}
 & $2x+y$
 \\ \cline{1-4}
\multirow{6}{*}{\texttt{imp.}}
 & \texttt{incr1}
 & \gape{\scalebox{0.7}{$f(x)=\begin{cases*}
      1+f(x+1) & if $x < 10$    \\[-2pt]
      1        & if $x \geq 10$ \\
    \end{cases*}$}}
 & $\max(1, 11-x)$
 \\ \cline{2-4}
 & \texttt{noisy\_strt1}
 & \gape{\scalebox{0.7}{$f(x)=\begin{cases*}
      0        & if $x = 0    \lor  x = 20$    \\[-2pt]
      1+f(x-1) & if $x \neq 0 \land x \neq 20$ \\
    \end{cases*}$}}
 & \gape{\scalebox{0.7}{$\begin{cases*}
      x    & if $x < 20$    \\[-2pt]
      x-20 & if $x \geq 20$ \\
    \end{cases*}$}}
 \\ \cline{2-4}
 & \texttt{noisy\_strt2}
 & \gape{\scalebox{0.7}{$f(x)=\begin{cases*}
      0        & if $x = 0    \lor  x = 65536$    \\[-2pt]
      1+f(x-1) & if $x \neq 0 \land x \neq 65536$ \\
    \end{cases*}$}}
 & \gape{\scalebox{0.7}{$\begin{cases*}
      x       & if $x < 65536$    \\[-2pt]
      x-65536 & if $x \geq 65536$ \\
    \end{cases*}$}}
 \\ \cline{2-4}
 & \texttt{multiphase1}
 & \gape{\scalebox{0.7}{$f(i,n,r)=\begin{cases*}
      0            & if $i \geq n$          \\[-2pt]
      1+f(0,n,r-1) & if $i < n \land r > 0$ \\[-2pt]
      1+f(i+1,n,r) & if $i < n \land r = 0$ \\
    \end{cases*}$}}
 & \gape{\scalebox{0.7}{$\begin{cases*}
      0   & if $i \geq n$          \\[-2pt]
      n+r & if $i < n \land r > 0$ \\[-2pt]
      n-i & if $i < n \land r = 0$ \\
    \end{cases*}$}}
 \\ \cline{2-4}
 & \texttt{lba\_ex\_viap}
 & \gape{\scalebox{0.7}{$f(x,y,c)=\begin{cases*}
      1+f(x+1,y+1,c) & if $x+y < c$    \\[-2pt]
      0              & if $x+y \geq c$ \\
    \end{cases*}$}}
 & \gape{\scalebox{0.7}{$\max\Big(0,\Big\lceil{\frac{c-(x+y)}{2}}\Big\rceil\Big)$}}
 \\ \cline{1-4} \pagebreak
\multirow{6.25}{*}{\texttt{nested}}
 & \texttt{nested}
 & \gape{\scalebox{0.7}{$f(x) =
    \begin{cases*}
      f(f(x-1)) + 1 & if $x > 0$ \\[-2pt]
      0             & if $x = 0$ \\
    \end{cases*}$}}
 & $x$
 \\ \cline{2-4}
 & \texttt{nested\_case}
 & \gape{\scalebox{0.7}{$f(x,b) =
    \begin{cases*}
      f(f(x-1,b),b) + 1 & if $x > 0 \land b=0$ \\[-2pt]
      x+f(x-1,b)        & if $x > 0 \land b>0$ \\[-2pt]
      0                 & if $x = 0$ \\
    \end{cases*}$}}
 & \gape{\scalebox{0.7}{$\begin{cases*}
      x                             & if $b = 0$ \\[-2pt]
      \frac{1}{2}x^2 + \frac{1}{2}x & if $b > 0$ \\[2pt]
    \end{cases*}$}}
 \\ \cline{2-4}
 & \texttt{nested\_div}
 & \gape{\scalebox{0.7}{$f(x) =
    \begin{cases*}
      f\Big(f\Big(\Big\lfloor\frac{x}{2}\Big\rfloor\Big)\Big) + 1 & if $x > 0$ \\
      0                                                           & if $x = 0$ \\
    \end{cases*}$}}
 & \gape{\scalebox{0.7}{$\begin{cases*}
      3 & if $x \geq 4$ \\[-2pt]
      2 & if $x = 3$    \\[-2pt]
      x & if $x \leq 2$ \\
    \end{cases*}$}}
 \\ \cline{2-4}
 & \texttt{mccarthy91}
 & \gape{\scalebox{0.7}{$f(x) =
    \begin{cases*}
      f(f(x+11)) & if $x \leq 100$ \\[-2pt]
      x-10       & if $x \geq 101$ \\
    \end{cases*}$}}
 & \gape{\scalebox{0.7}{$\begin{cases*}
      91   & if $x\leq 100$ \\[-2pt]
      x-10 & if $x\geq 101$ \\
    \end{cases*}$}}
 \\ \cline{2-4}
 & \texttt{golomb}
 & \gape{\scalebox{0.7}{$f(x) =
    \begin{cases*}
      f(x-f(x-1)) + 1 & if $x > 0$ \\[-2pt]
      1               & if $x = 0$ \\
    \end{cases*}$}}
 & \gape{$\Big\lfloor{\frac{1+\sqrt{8x}}{2}}\Big\rfloor$}
 \\ \cline{1-4} %
\multirow{8}{*}{\texttt{misc.}}
 & \texttt{div}
 & \gape{\scalebox{0.7}{$f(x,y) =
    \begin{cases*}
      f(x-y,y) + 1 & if $x \geq y \land y > 0$ \\[-2pt]
      0            & if $x < y    \land y > 0$   \\
    \end{cases*}$}}
 & $\Big\lfloor{\frac{x}{y}}\Big\rfloor$
 \\ \cline{2-4}
 & \texttt{sum-osc}
 & \gape{\scalebox{0.7}{$f(x,y) =
    \begin{cases*}
      f(x-1,y)   + 1 & if $x > 0 \land y > 0$ \\[-2pt]
      f(x+1,y-1) + y & if $x = 0 \land y > 0$ \\[-2pt]
      1              & if $y = 0$ \\
    \end{cases*}$}}
 & \gape{\scalebox{0.7}{$\begin{cases*}
      1                                 & if $y = 0$ \\[-2pt]
      \frac{1}{2}y^2 + \frac{3}{2}y + x & if $y > 0$ \\[2pt]
    \end{cases*}$}}
 \\ \cline{2-4}
 & \texttt{bin\_search}
 & \gape{\scalebox{0.7}{$f(x) =
    \begin{cases*}
      f\Big(\Big\lfloor\frac{x}{2}\Big\rfloor\Big) + 1 & if $x > 0$ \\
      0                                                & if $x = 0$ \\
    \end{cases*}$}}
 & \gape{\scalebox{0.7}{$\begin{cases*}
      0                                 & if $x = 0$ \\[-2pt]
      1+\big\lfloor\log_2(x)\big\rfloor & if $x > 0$
    \end{cases*}$}}
 \\ \cline{2-4}
 & \texttt{qsort\_best}
 & \gape{\scalebox{0.7}{$f(x) =
    \begin{cases*}
      f\Big(\Big\lfloor\frac{x-1}{2}\Big\rfloor\Big) + f\Big(\Big\lceil\frac{x-1}{2}\Big\rceil\Big) + x & if $x > 0$ \\
      1                                                                                                 & if $x = 0$ \\
    \end{cases*}$}}
 & \makecell[l]{
   \phantom{$\sim$ }$?$ \\
   $\sim x\mathrm{log}_2(x)$}
 \\ \cline{2-4}
 & \texttt{prs23\_1}
 & \gape{\makecell[l]{
     \scalebox{0.7}{$f(0)=1$}\\
     \scalebox{0.7}{$f(n+1)=\begin{cases*}
      2f(n) & if $f(n)<500$     \\[-2pt]
      f(n)  & if $f(n)\geq 500$ \\
     \end{cases*}$}
     \\[-2pt]
     \scalebox{0.7}{$g(0)=1$}\\
     \scalebox{0.7}{$g(n+1)=\begin{cases*}
      g(n)   & if $f(n)<500$                        \\[-2pt]
      g(n)+3 & if $f(n)\geq 500\land f(n)>g(n)$     \\[-2pt]
      g(n)-1 & if $f(n)\geq 500\land f(n)\leq g(n)$ \\
     \end{cases*}$}}}
 & \makecell[l]{
     \gape{\scalebox{0.7}{$f(n)=\begin{cases*}
      2^n & if $n < 9$    \\[-2pt]
      512 & if $n \geq 9$ \\
     \end{cases*}$}}
     \\
     \gape{\scalebox{0.7}{$g(n)=\begin{cases*}
      1                   & if $n < 9$          \\[-2pt]
      3n-26               & if $9 \geq n < 180$ \\[-2pt]
      514-(n\mathrm{\%}4) & if $n \geq 180$
    \end{cases*}$}}}
 \\ \cline{1-4}
\multirow{14}{*}{\texttt{CAS-style}}
 & \texttt{exp1}
 & \gape{\scalebox{0.7}{$f(n) =
    \begin{cases*}
      2f(n-1) & if $n > 0$ \\[-2pt]
      3       & if $n = 0$ \\
    \end{cases*}$}}
 & $3\times 2^n$
 \\ \cline{2-4}
 & \texttt{exp2}
 & \gape{\scalebox{0.7}{$f(n) =
    \begin{cases*}
      2f(n-1)+1 & if $n > 0$ \\[-2pt]
      3         & if $n = 0$ \\
    \end{cases*}$}}
 & $4\times 2^n - 1$
 \\ \cline{2-4}
 & \texttt{exp3}
 & \gape{\makecell[l]{
     \scalebox{0.7}{$f(x,y)=g(h(x,y))$}\\
     \scalebox{0.7}{$\phantom{,y}g(z)=\begin{cases*}
      2g(z-1) & if $z > 0$ \\[-2pt]
      1       & if $z = 0$ \\
     \end{cases*}$}
     \\[.75em]
     \scalebox{0.7}{$h(x,y)=\begin{cases*}
      h(x-1,y)+1 & if $x > 0$ \\[-2pt]
      y          & if $x = 0$ \\
     \end{cases*}$}}}
 & $f(x,y) = 2^{x+y}$
 \\ \cline{2-4}
 & \texttt{fib}
 & \gape{\scalebox{0.7}{$f(n) =
    \begin{cases*}
      f(n-1)+f(n-2) & if $n > 1$ \\[-2pt]
      1             & if $n = 1$ \\[-2pt]
      0             & if $n = 0$ \\
    \end{cases*}$}}
 & \gape{$\frac{1}{\sqrt{5}}\Big(\big(\frac{1+\sqrt{5}}{2}\big)^n-\big(\frac{1-\sqrt{5}}{2}\big)^n\Big)$}
 \\[-0.5em] \cline{2-4}
 & \texttt{harmonic}
 & \gape{\scalebox{0.7}{$f(n) =
    \begin{cases*}
      f(n-1)+\frac{1}{n} & if $n > 0$ \\[-2pt]
      0                  & if $n = 0$ \\
    \end{cases*}$}}
 & \makecell[l]{
   \phantom{$\sim$ }$\sum_{k=1}^n \frac{1}{k}$ \\
   $\sim \mathrm{ln}(n)$}
 \\ \cline{2-4}
 & \texttt{fact}
 & \gape{\scalebox{0.7}{$f(n) =
    \begin{cases*}
      nf(n-1) & if $n > 0$ \\[-2pt]
      1       & if $n = 0$ \\
    \end{cases*}$}}
 & \makecell[l]{
   \phantom{$=$ }$n!$ \\
   $=e^{\Theta(n \log n)}$}
 \\ \cline{2-4}
 & \texttt{cas\_st1}
 & \gape{\scalebox{0.7}{$f(n) =
    \begin{cases*}
      2nf(n-1) & if $n > 0$ \\[-2pt]
      1        & if $n = 0$ \\
    \end{cases*}$}}
 & \makecell[l]{
   \phantom{$=$ }$2^n n!$ \\
   $=e^{\Theta(n \log n)}$}
 \\ \cline{2-4}
 & \texttt{cas\_st2}
 & \gape{\scalebox{0.7}{$f(n) =
    \begin{cases*}
      n^2f(n-1) & if $n > 0$ \\[-2pt]
      1         & if $n = 0$ \\
    \end{cases*}$}}
 & \makecell[l]{
   \phantom{$=$ }$(n!)^2$ \\
   $=e^{\Theta(n \log n)}$}
 \\ \cline{2-4}
 & \texttt{cas\_st3}
 & \gape{\scalebox{0.7}{$f(n) =
    \begin{cases*}
      nf(n-1)+1 & if $n > 0$ \\[-2pt]
      1         & if $n = 0$ \\
    \end{cases*}$}}
 & \makecell[l]{
   \phantom{$\sim$ }$n! \sum_{k=0}^n \frac{1}{k!}$ \\
   $\sim e n!$}
 \\ \cline{2-4}
 & \texttt{cas\_st4}
 & \gape{\scalebox{0.7}{$f(n) =
    \begin{cases*}
      2f(n-1)+\frac{1}{n} & if $n > 0$ \\[-2pt]
      0                   & if $n = 0$ \\
    \end{cases*}$}}
 & \makecell[l]{
   \phantom{$\sim$ }$2^n \sum_{k=1}^n \frac{1}{2^k k}$ \\
   $\sim \mathrm{ln}(2) \times 2^n$}
 \\ \cline{2-4}
 & \texttt{cas\_st5}
 & \gape{\scalebox{0.7}{$f(n) =
    \begin{cases*}
      2nf(n-1)+\frac{1}{n} & if $n > 0$ \\[-2pt]
      0                    & if $n = 0$ \\
    \end{cases*}$}}
 & {\scriptsize\makecell[l]{
   \phantom{$\sim$ }$2^n n! \sum_{k=1}^n \frac{1}{k 2^k k!}$ \\
   $\sim (\mathrm{Ei}(1/2)+\mathrm{ln}(2)\!-\!\gamma)\cdot 2^n n!$}}
   \vspace*{-1em}
 \\ \cline{1-4}
\end{longtable}}

\clearpage
\newpage

\begin{table}[h!]
\centering
\caption{Experimental evaluation and comparison}
\label{table:newexp1}
{\footnotesize
 \setlength{\extrarowheight}{2.4pt}%
 \setlength{\tabcolsep}{5pt}%
 \vspace{-1em}
\begin{tabular}{c c | c c c | c | c c c c c c c c c c | c c c}
\hline
\textbf{Category} & \textbf{Bench}
& \rotatebox[origin=c]{270}{\texttt{mlsolve(L)}}
& \rotatebox[origin=c]{270}{\scriptsize\texttt{mlsolve(L)+domsplit}}
& \rotatebox[origin=c]{270}{\scriptsize\texttt{mlsolve(S)+domsplit}}
& \rotatebox[origin=c]{270}{\texttt{Ciao's builtin}}
& \rotatebox[origin=c]{270}{\texttt{RaML}}
& \rotatebox[origin=c]{270}{\texttt{PUBS}}
& \rotatebox[origin=c]{270}{\texttt{Cofloco}}
& \rotatebox[origin=c]{270}{\texttt{KoAT}}
& \rotatebox[origin=c]{270}{\texttt{Duet-ICRA18}}
& \rotatebox[origin=c]{270}{\texttt{Duet-CHORA}}
& \rotatebox[origin=c]{270}{\texttt{Duet-CRA23}}
& \rotatebox[origin=c]{270}{\texttt{Duet-CRA23 (*)}}
& \rotatebox[origin=c]{270}{\texttt{Loopus15 (*)}}
& \rotatebox[origin=c]{270}{\texttt{PRS23's solver}}
& \rotatebox[origin=c]{270}{\texttt{Sympy}}
& \rotatebox[origin=c]{270}{\texttt{PURRS}}
& \rotatebox[origin=c]{270}{\texttt{Mathematica}}
\\ \hline
\multirow{5}{*}{\texttt{scale}}
 &\texttt{highdim1}
 & \good & \good & \bad & \none
 & \good & \approxbnd & \good & \approxbnd
 & \none & \none & \none & \none & \good
 & \none & \none & \none & \none \\ \cline{2-19}
 &\texttt{poly2}
 & \good & \good & \good & \good
 & \good & \good & \good & \none
 & \none & \none & \none & \good & \good
 & \none & \none & \none & \good \\ \cline{2-19}
 &\texttt{poly5}
 & \good & \good & \good & \good
 & \good & \good & \good & \none
 & \none & \none & \none & \none & \good
 & \none & \none & \none & \good \\ \cline{2-19}
 &\texttt{poly7}
 & \bad & \bad & \good & \good
 & \good & \good & \good & \none
 & \none & \none & \none & \none & \good
 & \none & \none & \none & \good \\ \cline{2-19}
 &\texttt{highdim2}
 & \none & \none & \bad & \good
 & \none & \good & \good & \none
 & \none & \none & \none & \none & \bad
 & \none & \none & \none & \good \\ \cline{1-19}
\multirow{4}{*}{\texttt{amortized}}
 &\texttt{loop\_tarjan}
 & \good & \good & \good & \none
 & \good & \good & \good & \approxbnd
 & \good & \good & \good & \good & \approxbnd
 & \none & \none & \none & \none \\ \cline{2-19}
 &\texttt{enqdeq1}
 & \good & \good & \good & \none
 & \good & \none & \approxbnd & \none
 & \none & \none & \none & \good & \bad
 & \none & \none & \none & \none \\ \cline{2-19}
 &\texttt{enqdeq2}
 & \bad & \bad & \approxnobnd & \none
 & \approxbnd & \none & \approxbnd & \none
 & \none & \none & \none & \approxbnd & \bad
 & \none & \none & \none & \none \\ \cline{2-19}
 &\texttt{enqdeq3}
 & \good & \good & \good & \none
 & \approxbnd & \none & \approxbnd & \none
 & \none & \none & \none & \approxbnd & \bad
 & \none & \none & \none & \none \\ \cline{1-19}
\multirow{5}{*}{\texttt{max-heavy}}
 &\texttt{merge-sz}
 & \good & \good & \good & \good
 & \good & \approxbnd & \good & \approxbnd
 & \good & \good & \none & \good & \approxbnd
 & \none & \none & \none & \none \\ \cline{2-19}
 &\texttt{merge}
 & \bad & \good & \good & \bad
 & \bad & \good & \good & \bad
 & \bad & \bad & \none & \bad & \bad
 & \none & \none & \none & \none \\ \cline{2-19}
 &\texttt{open-zip}
 & \good & \good & \good & \none
 & \approxbnd & \approxbnd & \good & \approxbnd
 & \approxbnd & \approxbnd & \none & \none & \approxbnd
 & \none & \none & \none & \none \\ \cline{2-19}
 &\texttt{s-max}
 & \good & \good & \good & \approxbnd
 & \good & \approxbnd & \good & \approxbnd
 & \approxbnd & \good & \none & \good & \approxbnd
 & \none & \none & \none & \none \\ \cline{2-19}
 &\texttt{s-max-1}
 & \good & \good & \good & \none
 & \good & \good & \good & \approxbnd
 & \approxbnd & \good & \none & \approxbnd & \approxbnd
 & \none & \none & \none & \none \\ \cline{1-19}
\multirow{5}{*}{\texttt{imp.}}
 &\texttt{incr1}
 & \bad & \good & \good & \none
 & \none & \good & \good & \bad
 & \approxbnd & \good & \none & \good & \good
 & \none & \none & \none & \none \\ \cline{2-19}
 &\texttt{noisy\_strt1}
 & \none & \good & \good & \approxbnd
 & \approxbnd & \approxbnd & \good & \approxbnd
 & \approxbnd & \approxbnd & \approxbnd & \approxbnd & \approxnobnd
 & \good & \none & \approxnobnd & \approxnobnd \\ \cline{2-19}
 &\texttt{noisy\_strt2}
 & \approxbnd & \approxbnd & \approxbnd & \approxbnd
 & \approxbnd & \approxbnd & \good & \approxbnd
 & \approxbnd & \none & \approxbnd & \approxbnd & \approxnobnd
 & \good & \none & \approxnobnd & \approxnobnd \\ \cline{2-19}
 &\texttt{multiphase1}
 & \bad & \good & \good & \none
 & \none & \none & \good & \bad
 & \bad & \none & \bad & \bad & \bad
 & \none & \none & \none & \none \\ \cline{2-19}
 &\texttt{lba\_ex\_viap}
 & \none & \approxnobnd & \approxnobnd & \none
 & \none & \approxbnd & \approxnobnd & \bad
 & \approxbnd & \approxbnd & \approxbnd & \approxbnd & \approxbnd
 & \none & \none & \none & \none \\ \cline{1-19}
\multirow{5}{*}{\texttt{nested}}
 &\texttt{nested}
 & \good & \good & \good & \none
 & \good & \none & \none & \none
 & \approxbnd & \none & \none & \none & \none
 & \none & \none & \none & \none \\ \cline{2-19}
 &\texttt{nested\_case}
 & \bad & \good & \good & \none
 & \none & \none & \none & \none
 & \none & \none & \none & \none & \none
 & \none & \none & \none & \none \\ \cline{2-19}
 &\texttt{nested\_div}
 & \bad & \bad & \good & \none
 & \bad & \none & \none & \none
 & \none & \none & \none & \none & \none
 & \none & \none & \none & \none \\ \cline{2-19}
 &\texttt{mccarthy91}
 & \none & \bad & \good & \none
 & \none & \none & \bad & \none
 & \none & \none & \none & \none & \none
 & \none & \none & \none & \none \\ \cline{2-19}
 &\texttt{golomb}
 & \approxnobnd & \approxnobnd & \bad & \none
 & \bad & \none & \none & \none
 & \none & \none & \none & \none & \none
 & \none & \none & \none & \none \\ \cline{1-19}
\multirow{5}{*}{\texttt{misc.}}
 &\texttt{div}
 & \good & \good & \good & \none
 & \none & \bad & \bad & \bad
 & \bad & \bad & \none & \none & \bad
 & \none & \none & \none & \approxbnd \\ \cline{2-19}
 &\texttt{sum-osc}
 & \bad & \good & \approxnobnd & \none
 & \bad & \bad & \bad & \bad
 & \none & \none & \none & \none & \none
 & \none & \none & \none & \none \\ \cline{2-19}
 &\texttt{bin\_search}
 & \approxbnd & \good & \good & \none
 & \bad & \approxbnd & \none & \none
 & \none & \bad & \none & \bad & \bad
 & \none & \none & \approxbnd & \approxnobnd \\ \cline{2-19}
 &\texttt{qsort\_best}
 & \bad & \approxnobnd & \bad & \none %
 & \bad & \bad & \none & \none
 & \none & \bad & \none & \none & \none
 & \none & \none & \approxbnd & \approxbnd \\ \cline{2-19} %
 &\texttt{prs23\_1}
 & \none & \none & \none & \none
 & \none & \none & \none & \none
 & \none & \bad & \bad & \bad & \bad
 & \good & \none & \none & \none \\ \cline{1-19}
\multirow{11}{*}{\texttt{CAS-style}}
 &\texttt{exp1}
 & \good & \good & \good & \good
 & \none & \good & \none & \none
 & \good & \good & \none & \good & \none
 & \good & \good & \good & \good \\ \cline{2-19}
 &\texttt{exp2}
 & \good & \good & \approxbnd & \good
 & \none & \good & \none & \none
 & \good & \good & \none & \good & \none
 & \good & \good & \good & \good \\ \cline{2-19}
 &\texttt{exp3}
 & \bad & \bad & \good & \good
 & \none & \none & \none & \none
 & \good & \good & \none & \good & \none
 & \none & \none & \none & \good \\ \cline{2-19}
 &\texttt{fib}
 & \bad & \bad & {\scriptsize $e^{\theta}$} & \good
 & \none & {\scriptsize $e^\theta$} & \none & \none
 & \none & {\scriptsize $e^\theta$} & \none & \none & \none
 & \good & \good & \good & \good \\ \cline{2-19}
 &\texttt{harmonic}
 & \bad & \bad & \approxnobnd & \good
 & \none & \none & \none & \none
 & \none & \none & \none & \none & \none
 & \none & \none & \good & \good \\ \cline{2-19}
 &\texttt{fact}
 & \approxbnd & \approxbnd & \good & \good
 & \none & \none & \bad & \none
 & \none & \none & \none & \none & \none
 & \none & \good & \good & \good \\ \cline{2-19}
 &\texttt{cas\_st1}
 & \bad & \bad & \good & \good
 & \none & \none & \bad & \none
 & \none & \none & \none & \none & \none
 & \none & \good & \good & \good \\ \cline{2-19}
 &\texttt{cas\_st2}
 & {\scriptsize $e^{\theta}$} & {\scriptsize $e^{\theta}$} & \good & \good
 & \none & \none & \none & \none
 & \none & \none & \none & \none & \none
 & \none & \none & \good & \good \\ \cline{2-19}
 &\texttt{cas\_st3}
 & \approxbnd & \approxbnd & \approxnobnd & \good
 & \none & \none & \none & \none
 & \none & \none & \none & \none & \none
 & \none & \none & \good & \good \\ \cline{2-19}
 &\texttt{cas\_st4}
 & \approxnobnd & \approxnobnd & \approxbnd & \good
 & \none & \none & \none & \none
 & \none & \none & \none & \none & \none
 & \none & \none & \good & \good \\ \cline{2-19}
 &\texttt{cas\_st5}
 & {\scriptsize $e^{\theta}$} & {\scriptsize $e^{\theta}$} & \approxbnd & \good
 & \none & \none & \none & \none
 & \none & \none & \none & \none & \none
 & \none & \none & \good & \good \\[\extrarowheight]\cline{1-19}%
 \multicolumn{2}{c|}{\vspace{-\extrarowheight}Total number of \good\;(/40)}
 & $14$ & $21$ & $25$ & $16$
 & $10$ & $10$ & $15$ & $0$
 & $5$ & $8$ & $1$ & $9$ & $5$
 & $6$ & $5$ & $10$ & $15$
\\[-\extrarowheight] \hline
\end{tabular}}
\end{table}

\clearpage
\newpage

\paragraph[Experimental setup]{\textbf{Experimental setup}.}
Chosen hyperparameters are given for the sake of completeness.

For regularized linear regression (Algorithm~\ref{algo1}), we set $\epsilon =
0.05$ for feature selection, $k=2$ for cross-validation, which will be performed
to choose the value of $\lambda$ in a set $\Lambda$ of $100$ values taken from
the interval $[0.001, 1]$. (It is also possible to construct the entire
regularization path for the lasso, which is piecewise linear, and obtain the
solution for every value of $\lambda \in \rr$ \citep{Hastie-15a}, but here we
opted to provide an explicit set of possible values $\Lambda$.)
If we use domain splitting, the choice of regression
domain is as described in
Section~\ref{sec:description-approach-domain-splitting}
(fit separately on the subdomains corresponding to each clause), but if we
do not, we choose $\dd$ to be (a power of) $\naturalnumbers_{>0}$ (this
avoids the pathological behavior in $0$ of some candidates). Verification is
still performed on the whole domain of the equation, i.e., $\varphi_{pre}=\vee_i \varphi_i$.
For each run of Algorithm~\ref{algo1}, $\numtraining=100$ input values in $\dd$ are
generated, with each variable in $[0,\bndtraining]$, where $\bndtraining\in\{20,10,5,3\}$ is chosen so %
that generation ends within a timeout of $2$ seconds.\footnote{This
\texttt{Python} prototype does not use tabling, and evaluation can be long for
recurrences with non-linear recursion. Much higher values of $\bndtraining$ could be chosen
with a less naive evaluation strategy.}

Candidate functions are selected as follows. For each number $\numparam$ of input
variables we choose sets $\basefuns_{\mathrm{small}}$, $\basefuns_{\mathrm{medium}}$,
$\basefuns_{\mathrm{large}}$ representative of common complexities encountered in cost
analysis. They are chosen explicitly for $\numparam=1$ and $2$, and built as
combinations of base functions for $\numparam\geq 3$. Regression is performed for each
set, with a timeout of $10$ seconds,\footnote{For each subdomain in the case of \texttt{domsplit}.}
and the best solution is chosen.
Chosen sets are indicated in Figures~\ref{fig:candidates-smalldim}
and \ref{fig:candidates-highdim}.

\begin{figure}[h!]
\footnotesize
{\setlength{\fboxsep}{0pt}\framebox{
\begin{minipage}{\linewidth}
\[\begin{array}{cccc}
 \numparam & S_{\mathrm{small}} & S_{\mathrm{medium}} & S_{\mathrm{large}} \\
 \hline
 1 & \{x, x^2\}
   & \{\lceil\log_2(x)\rceil, \lfloor\sqrt{x}\rfloor, x\lceil\log_2(x)\rceil\}
   & \{\lfloor\log_2(x)\rfloor, x\lfloor\log_2(x)\rfloor, 2^x, 5^x, x\cdot2^x, x!\}\\[3pt]
 2 & \{x, y\}
   & \{x^2,xy,y^2,x^2y,xy^2,x^2y^2\}
   & \left\{\begin{array}{l}
       \lfloor\frac{x}{y}\rfloor,\lfloor\frac{y}{x}\rfloor, \lceil\frac{x}{y}\rceil, \lceil\frac{y}{x}\rceil,
       2^x, 2^y, \max(x,y),\\
       \lceil\log_2(x)\rceil, \lfloor\log_2(x)\rfloor, x\lceil\log_2(x)\rceil, x\lfloor\log_2(x)\rfloor,\\
       \lceil\log_2(y)\rceil, \lfloor\log_2(y)\rfloor, y\lceil\log_2(y)\rceil, y\lfloor\log_2(y)\rfloor,\\
     \end{array}\right\}\\
\end{array}\]
\vspace{3pt}
\end{minipage}
}}
\vspace{-1em}
\caption{Candidate functions for linear regression in dimension $\leq 2$.
  ${\basefuns_{\mathrm{small}}=S_{\mathrm{small}}}$,
  ${\basefuns_{\mathrm{medium}}=\basefuns_{\mathrm{small}}\cup S_{\mathrm{medium}}}$,
  ${\basefuns_{\mathrm{large}}=\basefuns_{\mathrm{medium}}\cup S_{\mathrm{large}}}$. Lambdas omitted for conciseness.}
\label{fig:candidates-smalldim}
\framebox{
\begin{minipage}{\linewidth}
$\tilde{\basefuns}_{\mathrm{small}}=\{x\}$,
$\tilde{\basefuns}_{\mathrm{medium}}=\{x,x^2\}$,
$\tilde{\basefuns}_{\mathrm{large}}=\{x,x^2,x^3\}$.
For $s\in\{\mathrm{small},\mathrm{medium},\mathrm{large}\}$,
\begin{equation*}\small
 \basefuns_{s}=\bigcup_{\substack{k\in\nn\\k\leq \numparam+|\tilde{\basefuns}_s|}}
 \Big\{\prod_{i=1}^k f_i(x_{\selcomb(i)}) \,\Big|\,
       \forall i\, f_i\in\tilde{\basefuns}_s,\,
       \selcomb : [1,k] \to [1,\numparam],\,
       \forall i,j\, \big(i\neq j \implies (f_i,\selcomb(i))\neq(f_j,\selcomb(j))\big)\Big\}
\end{equation*}
\end{minipage}
}
\vspace{-1em}
\caption{Candidate functions for linear regression in dimension $\geq 3$.
$\basefuns_{s}$ is defined by combination of simpler base functions,
as bounded products of applications of base functions to individual
input variables.
For example, with $\numparam=3$, $\basefuns_{\text{small}}=\{x,y,z,xy,xz,yx,xyz\}$.
}
\label{fig:candidates-highdim}
\end{figure}

In the case of symbolic regression, instead of candidates, we need to choose
operators allowed in expression trees. We choose binary operators $\{+, -, \max,
\times, \div, (\cdot)^{(\cdot)}\}$ and unary operators $\{\lfloor\cdot\rfloor,
\lceil\cdot\rceil, (\cdot)^2, (\cdot)^3, \log_2(\cdot), 2^{(\cdot)},
(\cdot)!\}$. Nodes, including leaves, have default cost $1$. Ceil and floor have
cost $2$ and binary exponentiation has cost $3$. We have $45$ populations of
$33$ individuals, and run \texttt{PySR} for $40$ iterations and otherwise
default options.

In all cases, to account for randomness in the algorithms, experiments are run
twice and the best solution is kept. Variability is very low when exact
solutions can be found.

Experiments are run on a small Linux laptop with
\texttt{1.1GHz Intel Celeron N4500} CPU,
\texttt{4 GB}, \texttt{2933 MHz DDR4} memory.
Lasso regression takes 1.4s to 8.4s with mean 2.5s (on each candidate
set, each benchmark, excluding equation evaluation and regression
timeouts\footnote{Timeouts occur for all base function sets on only
$1$ of $40$ benchmarks, \texttt{highdim2}. They also occur on some
functions sets for 2 additional benchmarks, \texttt{highdim1}
(``medium'' and ``large'' sets) and \texttt{multiphase1} (``large''
base function set).}),
showing
that our approach is reasonably efficient. 
Symbolic regression (on each subdomain, no timeout, without early exit
conditions) takes 56s to 133s with mean 66s.
It may be noted that these results correspond to initial prototypes,
and that experiments focus on testing the accuracy of our approach:
many avenues for easy optimizations are left open, e.g., stricter
(genetic) resource management in symbolic regression.

\paragraph[Benchmarks]{\textbf{Benchmarks}.}

We present 40 benchmarks, i.e., (systems of) recurrence equations, extracted
from our experiments and organized in 7 categories that test different features.
We believe that such benchmarks, evenly distributed among the
categories considered, are
reasonably representative and capture the essence of a much larger set
of benchmarks, including those that arise in static cost analysis,
and that their size and diversity provide experimental insight on the
scalability of our approach.

Category \texttt{scale} tests how well solvers can handle increasing dimensions
of the input space, for simple linear or polynomial solutions. The intuition for
this test is that (global and numerical) machine learning methods may suffer
more from the curse of dimensionality than (compositional and symbolic)
classical approaches. We also want to evaluate when the curse will start to
become an issue.
Here, functions are multivariate, and most benchmarks are systems of
equations, in order to easily represent nested loops giving rise to polynomial cost.

Category \texttt{amortized} contains examples inspired by the amortized resource
analysis line of work illustrated by \texttt{RaML}. Such problems may be seen as
situations in which the worst-case cost of a function is not immediately given
by the sum of worst-case costs of called functions, requiring some information
on intermediate state of data structures, which is often tracked using a notion
of \emph{potential} \citep{tarjan85}. For the sake of comparison, we also track
those intermediate states as \emph{sizes}, and encode the corresponding
benchmarks in an equivalent way as systems of equations on \emph{cost} and
\emph{size}, following an approach
based\footnote{\label{footnote:amortized-eq-encoding}In~\cite{caslog-short}, the
authors set up the equations at the same time as they reduce them using
intermediate approximate solutions. Here, we are referring to the full system of
size/cost equations representing the program, without
simplifications/reductions.} on~\citep{caslog-short}. \texttt{loop\_tarjan}
comes from \citep{Sinn17-loopus}, whereas \texttt{enqdeq1-3} are adapted from
\citep{jan-hoffman-phd}. In each example, the cost function of the entry
predicate admits a simple representation, whereas the exact solutions for
intermediate functions are more complicated.

\begin{figure}[]
  \centering
  \prettylstciao
\vspace{-1em}
\begin{tabular}{p{6.5cm} p{5.5cm}}
\begin{lstlisting}[basicstyle=\scriptsize\ttfamily, linewidth=6cm]
enq((Li, Lo), X, ([X|Li], Lo)) :-
    tick(1).

deq((Li, [X|Lo]), X, (Li, Lo)) :-
    tick(1).
deq(([], []),    _X, ([], [])) :-
    tick(1).
deq((Li, []),     X, ([], Lo)) :-
    length(Li, N), tick(N),
    reverse(Li, [X|Lo]),
    tick(1).
\end{lstlisting}&
\begin{lstlisting}[basicstyle=\scriptsize\ttfamily, firstnumber=14, linewidth=5cm]
menq(Q, 0, Q).
menq(Q, N, QR) :- N>0,
    N1 is N-1,
    enq(Q, _, Q1),
    menq(Q1, N1, QR).

mdeq(Q, 0, Q).
mdeq(Q, N, QR) :- N>0,
    N1 is N-1,
    deq(Q, _, Q1),
    mdeq(Q1, N1, QR).
\end{lstlisting}
 \end{tabular}\vspace{-2em}
\begin{tabular}{c}
\begin{lstlisting}[basicstyle=\scriptsize\ttfamily, firstnumber=27, linewidth=6.5cm]
enqdeq1(N) :-
    menq(([],[]), N, Q), mdeq(Q, N, _).
enqdeq2(N, M, K) :- length(Lo, K),
    menq(([],Lo), N, Q), mdeq(Q, M, _).
enqdeq3(N) :- N2 is 2*N,
    menq(([],[]), N2, Q1),
    mdeq(Q1, N, Q2),
    menq(Q2, N, Q3),
    mdeq(Q3, N, _).
\end{lstlisting}
\end{tabular}\vspace{-.25em}
  \caption{\small Prolog encoding of the \texttt{enqdeq} benchmarks,
  inspired from Section~2.1 of \cite{jan-hoffman-phd}
  introducing amortized analysis,
  where a queue datastructure is implemented as two lists that act as stacks.
  We encode the \texttt{enqdeq} problems for each tool following a best-effort approach.
  Recurrence equations are set up in terms of compositions of cost and size functions.}
  \label{fig:enqdeq}
\end{figure}

Category \texttt{max-heavy} contains examples where either the equation or (the
most natural expression of) the solution contain $\max$ operators. Such
equations may typically arise in the worst-case analysis of non-deterministic
programs (or complex deterministic programs simplified to simple
non-deterministic programs during analysis), but may also appear naturally in
size equations of programs with tests.

Category \texttt{imp.} tests features crucial to the analysis of imperative
programs, that tend to be of lesser criticality in other paradigms. Those
features can be seen as features of loops (described as challenges in recent
work including \cite{montoya-phdthesis} and \cite{Sinn17-loopus}), such as
increasing variables (corresponds to recursive call with increasing arguments),
noisy/exceptional behavior (e.g., early exit), resets and multiphase loops.

Category \texttt{nested} contains examples with nested recursion, a class of
equations known for its difficulty \citep{tanny-talk-nestedreceq}, that can even
be used to prove undecidability of existence of solutions for recurrence
equations \citep{nested-receq-undecidable}. Such equations are extremely
challenging for ``reasoning-based'' approaches that focus on the structure of
the equation itself. However, focusing on the solution rather than the problem
(to put it another way, on semantics rather than syntax), we may notice that
several of these equations admit simple solutions, that may hence be guessed via
regression or template methods. We include the famous McCarthy 91 function
\citep{mccarthy91-original70}.

Category \texttt{misc.} features examples with arithmetic (euclidean
division), sublinear (logarithmic) growth, deterministic
divide-and-conquer, and alternating activation of equations' clauses.
We are not aware of simple closed-forms for the solution of
\texttt{qsort\_best}, although its asymptotic behavior is known.
Hence, we do not expect tools to be able to infer exact solutions, but
rather to aim for good approximations and/or bounds.
Example \texttt{prs23\_1} is given as a motivating example in
\cite{Wang-PRS23} for studying eventually periodic conditional linear
recurrences, and displays a particular difficulty: cases are defined by
conditions on results of recursive calls rather than simply on input values.

Finally, category \texttt{CAS-style} contains several examples from CAS's
problem domain, i.e., that can be reduced to unconditional single variable
equations with a single recursive case of shape $f(n) = a(n)\times f(n-1) +
b(n)$, with $a$ and $b$ potentially complicated, e.g., hypergeometric. This is
quite different from the application case of program cost analysis, where we
have multiple variables, conditionals are relevant, but recursive cases are
closer to (combinations of) $f(n) = a\times f(\phi(n)) + b(n)$, with $a$
constant but $\phi(n)$ non-trivial.\vspace{1em}

In the case of a system of equations, we only evaluate the solution found for
the topmost predicate (the entry function).
It may be noted that all of our equations admit a single solution $f:\dd\to\realnumbers$
where $\dd\subset\naturalnumbers^\numparam$ is defined by
$\varphi_{pre}=\vee_i\varphi_i$, and that the evaluation strategy (defined in
Section~\ref{sec:prelim}) terminates everywhere on $\dd$.
Precise handling and evaluation of partial solutions and non-deterministic
equations is left for future work.

\paragraph[Tools and benchmark translation]{\textbf{Tools and benchmark translation}.}

Obtaining meaningful comparisons of static analysis tools is a challenge.
Different tools analyze programs written in different languages, discrepancies
from algorithms to implementations are difficult to avoid, and tools are
particularly sensitive to representation choices: multiple programs, resp.,
equations, may have the same (abstract) semantics, resp., solutions, but be far
from equally easy to analyze.
In order to compare tools anyway, a common choice is to use off-the-shelf
translators to target each input language, despite inevitable and unequal
accuracy losses caused by those automated transformations.
In this paper, we adopt a different approach: each benchmark is written
separately in the input format of each tool, aiming for representations adapted
to each of them, following a best-effort approach.
This is motivated by the lack of available translation tools, by the loss of
accuracy observed for those we could find, and by the desire to nonetheless
offer a broad comparison with related work.

We cannot list all choices made in the translation process, and only present the
most important ones. As a guiding principle, we preserve as much as possible the
underlying numerical constraints and control-flow defined by each benchmark,
whether the problem is being encoded as a program or a recurrence equation.
In principle, it should be straightforward to translate between these
representations: a system of recurrence equations may directly be written as a
numerical program with recursive functions. However, there is an obstruction,
caused by missing features in the tools we evaluated. When it is sufficiently
easy to rewrite a benchmark in order to avoid the missing feature, we do so, but
otherwise we have to deem the problem unsupported.

Main missing features include support for systems of equations, multiprocedural
programs, multivariate equations, loop structures beyond a single incremented
index variable, non-determinism, complex numerical operations (polynomials,
divisions, rounding functions, but also maximum and minimum), and finally
support for general recursion, including nested calls and non-linear recursion.

For tools focusing on loops, with limited support for multiprocedural programs,
we perform the mechanical translation of linear tail-recursion to loops when
possible, even if this might be seen as a non-trivial help. Such cases are
indicated by ``\texttt{(*)}'' in our table. This includes \texttt{Loopus}, which
analyzes C and LLVM programs without recursive functions, and
\texttt{Duet-CRA23}, i.e., the 2023 version of \texttt{duet}'s \texttt{cra}
analysis.\footnote{Main branch, tested as
\url{https://github.com/zkincaid/duet/tree/7a5bb0fad9}, May 25, 2023.} Since
\texttt{Duet-CRA23} is still able to perform some interprocedural analysis, we
report its results on recursive-style benchmarks (column without
``\texttt{(*)}''), but it gives better results when this imperative translation
is applied.\footnote{\texttt{Duet}'s README indicates that the discontinued
feature ICRA for analysis of general recursive programs may still be found in
the \texttt{Newton-ark2} branch. Unfortunately, we were neither able to get it
(\texttt{008d278f52}, May 2020) to run, nor the most recent artifact with this
feature \citep{kincaid-closed-forms-popl19}, so we resort to the previous one
\citep{kincaid2018}, in Column \texttt{Duet-ICRA18}. Recursive functions are
also supported by a more template-based approach \citep{Breck-CHORA}, tested in
column \texttt{Duet-CHORA}.} We sometimes attempt less mechanical
transformations to loops, e.g., for \texttt{fib}. The transformation discussed
in this paragraph could also have been applied to \texttt{KoAT},\footnote{We
additionally would have needed to delay cost evaluation to the end, working on
size equations instead. It does improve the results in the case where
\texttt{KoAT} is limited by its lack of support for recursivity (discontinued
\texttt{Com\_2} construct), e.g., for \texttt{exp1} or
\texttt{fib}.}\textsuperscript{,}\footnote{KoAT version tested:
\url{https://github.com/aprove-developers/KoAT2-Releases/tree/c1c5290109}, May
10, 2023. ITS inputs, default options \texttt{koat2 analyse -i <filename>}.}
but we do not report full results for conciseness.

When support for explicit $\max$ operators is limited or absent, we transform
deterministic programs with $\max$ into non-deterministic programs, in a way that
preserves upper bounds, but in general not lower bounds.\footnote{For example, a
recursive call \texttt{return $\max(f(x-1,y), f(x,y-1))$} may be replaced by
\texttt{if(nondet()) return $f(x-1,y)$ else return $f(x,y-1)$}.} This transformation
is not needed for \texttt{mlsolve}, which accepts any nested expression with $\max$
and $\min$, but is required for \texttt{Ciao},\footnote{Several size and cost
analyses are available in Ciao. Here, we report on the one implemented in
\texttt{infercost}, available via \texttt{analyze(steps\_ualb)}, run on
numerical programs encoding the equations.} \texttt{PUBS}, \texttt{Cofloco},
\texttt{Loopus} and all versions of \texttt{duet}. It is applied for
\texttt{RaML} too, using a trick, as we did not find a native construct for
non-determinism. Finally, this cannot be applied or does not help to obtain
results for \texttt{PRS23}, \texttt{Sympy}, \texttt{PURRS} and
\texttt{Mathematica}.

\paragraph[Evaluation and Discussion]{\textbf{Evaluation and Discussion}.}

Overall, the experimental results are quite promising, showing that
our approach is accurate and efficient, as well as scalable. We first discuss the stacked
horizontal bar chart in Figure~\ref{fig:barchart}, which offers a
graphical and comprehensive overview of the accuracy of our approach
in comparison to state-of-the-art tools. We then provide comments on
Table~\ref{table:newexp1}, which presents more detailed information.

As mentioned earlier, bars \texttt{a}-\texttt{c} (the first three from
the top) of Figure~\ref{fig:barchart} represent the three implemented
versions of our approach. We can see that in general, for the
benchmark set considered, which we believe is reasonably
representative, each of these versions individually outperforms any of
the other state-of-the-art solvers (included in either cost analyzers
or CASs), which are represented by bars \texttt{d}-\texttt{q}.  Bar
\texttt{b} $+$ \texttt{c} represents the full version of our approach,
which combines both regression methods, lasso (\texttt{b}) and
symbolic (\texttt{c}), and the domain splitting strategy.
The three bars at the bottom of the chart are included to assess how
the combination of our machine learning-based solver with other
solvers would potentially improve the accuracy of the closed-forms
obtained. The idea is to use it as a complement of other back-end
solvers in a higher-level combined solver in order to obtain further
accuracy improvements.  We can see that the combination (bar
\texttt{b} $+$ \texttt{c} $+$ \texttt{d}) of \ciaopp's builtin solver
with the full version of our approach mentioned above
would obtain exact solutions for $85\%$ of the
benchmarks, and accurate approximations for the rest of them, except
for one benchmark. As shown in Table~\ref{table:newexp1}, this
benchmark is \texttt{prs23\_1}, which can only be solved by
\texttt{PRS23}'s solver. For our set of benchmarks, the best overall
accuracy would be achieved by the combination represented by the
bottom bar, which adds \texttt{PRS23}'s solver to the combination
above.

We now comment on the experimental results of Table~\ref{table:newexp1},
category by category.

For category \texttt{scale}, we notice that \texttt{mlsolve} can go up to
dimension $4$ without much trouble (actually up to $7$, i.e., querying for $f_4$
in \texttt{highdim2}, and failing after that). The impact of the curse of
dimensionality on our machine learning approach thus appears tolerable, and
comparable to that on the template-based \texttt{RaML}, which timeouts reaching
dimension $7$, i.e., for $f_4$ in \texttt{highdim2}, allowing
potentials to be polynomials of degree $7$. \texttt{PUBS} and \texttt{Cofloco}
perform well and quickly in this category, showing the value of compositional
and symbolic methods. Surprisingly, \texttt{Loopus} performs well only until
dimension $10$ where an approximation is performed, and \texttt{duet} gives
nearly no non-trivial answer. The benchmarks being naturally presented in a
multivariate recursive or nested way, they are difficult to encode in
\texttt{PRS23}, \texttt{Sympy} and \texttt{PURRS}. The polynomial (locally
monovariate) benchmarks are solved by \texttt{Mathematica}, if we provide the
order in which subequations must be solved. The benchmarks are similarly
difficult to encode in (cost-based) \texttt{KoAT}, and helping it with
manual control-flow linearization did not suffice to get ``\approxbnd''
results.

For category \texttt{amortized}, \texttt{RaML} performs well, but perhaps not as
well as could have been imagined. For \texttt{enqdeq}1,2,3, it outputs bounds
$[2n,3n]$, $[n+m,2n+m]$ and $[5n,8n]$, respectively. The simplicity of the
solutions allows \texttt{mlsolve} to perform well (using the full size and cost
recurrence equation encoding discussed above, cf. \cite{caslog-short},
Footnote~\ref{footnote:amortized-eq-encoding} and Figure~\ref{fig:enqdeq}). This
performance can be attributed to the fact that \texttt{mlsolve} does not need to
find and use the complicated cost expressions of the intermediate predicates:
only full execution traces starting from the entry point are considered. Note
that \texttt{Cofloco} also obtains reasonable results, making use of its
disjunctive analysis features (visible with the \texttt{-conditional\_ubs}
option), although it loses
accuracy in several places, and outputs linear
bounds with incorrect constants. After simplification of its output,
\texttt{Duet-CRA23} obtains $3n$, $2n+m$ and $8n$ on \texttt{enqdeq}, but it may
be noted that writing such benchmarks in an imperative fashion may be seen as
unreasonable help to the control-flow analysis. \texttt{Loopus} obtains only
quadratic bounds on the \texttt{enqdeq}, and no other tool obtains non-trivial
bounds. \texttt{loop\_tarjan} is handled by nearly all code analysis tools,
although surprisingly \texttt{Loopus} loses a minor amount
accuracy and outputs
$2n+1$ instead of $2n$. It may be noted that no tool gets the exact solution for
\texttt{enqdeq2}. For \texttt{mlsolve}, this hints at the fact that improved
domain splitting strategies may be beneficial, as they may allow it to discover the
$m\leq k$ constraint, which is not directly visible in the equations.

For category \texttt{max-heavy}, we simply observe that support for $\max$
operators is rare, although some disjunctive analysis and specific reasoning can
be enough to get exact results after the $\max\to\texttt{nondet}$
transformation, as is the case for \texttt{Cofloco}. Many tools can infer linear
bounds for these benchmarks, without being able to get the correct coefficients.
More challenging benchmarks could be created using nested $\min$-$\max$
expressions and non-linear solutions to separate more tools.

In category \texttt{imp.}, where examples are inspired from imperative problems,
we notice that the \texttt{domsplit} extension is crucial to allow
\texttt{mlsolve} to deal with even basic \texttt{for} loops (benchmark
\texttt{incr1}, which also encodes the classical difficulty of recurrence-based
methods with increasing arguments). \texttt{Cofloco} performs best in this
category thanks to its control-flow refinement features. For
\texttt{noisy\_strt}, several tools notice the two regimes but struggle to
combine them correctly and precisely. \texttt{PRS23} gets the exact solution, a
testimony to its precise (but monovariate) phase separation. As expected,
\texttt{mlsolve} fails on \texttt{noisy\_strt2} even when it succeeds on
\texttt{noisy\_strt1}, misled by the late change of behavior, invisible to naive
input generation, although counter-examples to the $x$ solution are produced.
Such counter-examples may be used in future work to guide the regression process
and learn better subdomains. Given the known limitations of \texttt{RaML} on
integers, some inequalities that are natural in an imperative setting are
difficult to encode without
accuracy loss. For tools that solve
\texttt{lba\_ex\_viap},\footnote{Inspired from an example in the \texttt{VIAP}
repository, \url{https://github.com/VerifierIntegerAssignment/}.} small
rounding errors appear.

Category \texttt{nested} is composed of particularly challenging problems for
which no general solving method is known, although low (model) complexity solutions
exist for particular cases, giving a strong advantage to the \texttt{mlsolve}
approach. It may be noted that \texttt{RaML} also gets the exact solution for
the \texttt{nested} benchmark and that \texttt{Duet-ICRA18} obtains the nearly
exact $\max(1,x)$ solution. \texttt{mlsolve(S)+domsplit} performs best. It
sometimes uses non-obvious expressions to fit the solution, such as
$\mathrm{ceil}(2.75 – 2.7/x)$ for the $x>0$ subdomain on \texttt{nested\_div}.
Although no tool solves the \texttt{golomb} recurrence in our experimental
setup, it may be noted that \texttt{mlsolve(S)} finds the exact solution when
$\sqrt{\cdot}$ is added to the list of allowed operators, showing the
flexibility of the approach.

In category \texttt{misc.}, division, logarithm and complex
control-flow, which are typically difficult to support, give an advantage to the
\texttt{mlsolve} approach, as control-flow has low impact and operators can
easily be included. It must be noted that our definition of ``precise
approximation'' makes linear bounds insufficient for the division and logarithm
benchmarks, and forces the $f(x,0)=1$ line to be considered in \texttt{sum-osc}.
For \texttt{bin\_search}, both \texttt{PUBS} and \texttt{PURRS} are able to obtain
$1+\log_2(x)$ bounds on the $x>0$ subdomain, close to the exact solution
$1+\lfloor\log_2(x)\rfloor$.
Closed-form solutions are not available for \texttt{qsort\_best}, and
the $x\log_2(x)$ asymptotic behavior is only recovered by
\texttt{mlsolve(L)+domsplit},\footnote{It may be noted that our chosen
hyperparameters, with a bound $\bndtraining=20$ on inputs, make it hard for
\texttt{mlcost} to distinguish base functions on such one-dimensional
benchmark. Results improve further if larger inputs are included in the
training set, as shown in \ref{appendix:full-output}.}
as well as \texttt{PURRS} and
\texttt{Mathematica} if helped by preprocessing, although other tools
do infer quadratic bounds.
Interestingly, \texttt{prs23\_1} is only solved by
\texttt{PRS23} \citep{Wang-PRS23}, because of the particular difficulty of using
recursive calls in update conditions, even though this naturally encodes
imperative loops with conditional updates.

Finally, as could be expected, \texttt{CAS-style} is the only category in which
computer algebra systems obtain good results. All monovariate benchmarks are
solved by \texttt{PURRS} and \texttt{Mathematica}, and only \texttt{Mathematica}
solves \texttt{exp3} (with hints on resolution order). Surprisingly,
\texttt{Sympy} fails for \texttt{harmonic} and \texttt{cas\_st2}, but still
generally obtains good results. It may be noted that for the last three
benchmarks, \texttt{Mathematica} represents solutions as special functions while
\texttt{PURRS} and \texttt{Ciao}'s builtin solver choose summations. These
representations are equivalent and none can be obviously preferred to the other.
The category is challenging for code analysis tools, but several deal or partly
deal with the case of exponential bounds and of \texttt{fib}. In this last case,
the exponential part of the bound obtained is $2^x$ for \texttt{PUBS} and
\texttt{Duet-CHORA} and $1.6169^x$ for \texttt{mlsolve(S)}, closer to the golden
ratio ($\varphi\approx 1.6180...$) expected at infinity. Interestingly, while
the exact solutions of the last \texttt{cas\_st*} are too complex to be
expressed by \texttt{mlsolve(S)} in our experimental setting, it is still able to
obtain very good approximations of the asymptotic behavior, returning $x!
\times 2.71828$, $2^x \times 0.69315$ and $2^x\times x! \times 0.5701515$ for
\texttt{cas\_st}3,4,5 respectively, with coefficients close to the constants $e
\approx 2.71828...$, $\ln(2)\approx 0.693147...$ and
$\mathrm{Ei}(1/2)+\mathrm{ln}(2)\!-\!\gamma \approx 0.57015142...$, showing its
ability to get good approximations, even in the case of complex equations with
complex solutions.\\

The results presented in \texttt{mlsolve} columns correspond to the
accuracy of proposed candidates.
We now say a few words on their verification.
It must be noted that while the \texttt{mlsolve} version presented in
Section~\ref{sec:description-approach} easily handles regression for systems of
equations, verification has only been implemented for single equations. The
verification strategy can be extended to systems of equations, but the naive
approach has the disadvantage of requiring candidates for all subfunctions. More
subtle solutions are left for future work.

Among the 40 benchmarks, \texttt{mlsolve(L)} finds the exact solution for 14, 9
of which come from single function problems. It is able to prove correctness of
all of those 9 candidates, using the translation described in
Equation~\ref{smtrep}.

As mentioned at the end of
Section~\ref{sec:description-approach-domain-splitting}, verification can be
performed in a similar way in the \texttt{domsplit} cases. This translation has
not yet been fully integrated in our prototype, but has been tested manually on
a few examples, showing that \texttt{Z3} can provide valuable insights for those
piecewise candidates (verification of the \texttt{noisy\_start1} candidate,
counter-examples, etc.), but is not strong enough to directly work with complex
expressions involving factorials, logarithms, or difficult exponents.

\paragraph[Additional experiment: comparison with symbolic (linear) equation solving]
{\textbf{Additional experiment: comparison with symbolic (linear) equation solving}.}

To further
justify our choice of methods, we conducted an additional experiment
comparing numerical linear regression (used in the case of \texttt{mlsolve(L)})
with symbolic linear equation solving in terms of efficiency.

As mentioned in the introduction, efficiency is not the primary reason
we choose search and optimization algorithms typical of machine
learning methods over more classical exact symbolic methods. Instead,
we prioritize expressivity and flexibility.
While symbolic linear equation solving is limited to situations where
an exact solution can be expressed as an affine combination of base
functions, our approach enables the discovery of approximate solutions
(particularly valuable in certain cost analysis applications) when
closed-form solutions are too complex to find.
Moreover, our modular approach allows the exploration of model spaces
more complex than affine ones, which cannot be handled by complete
methods (e.g., compare \texttt{mlsolve(S)} with \texttt{mlsolve(L)}).

Nevertheless, in situations where an exact solution can be expressed
as an affine combination of base functions, we may choose to use
\emph{exact, symbolic} linear equation solvers instead of
(\texttt{float}-based, iterative) linear regression solvers to infer
template coefficients, as we have done in this paper.
While this alternative approach is limited to cases where an exact
solution exists within the model space, it has the advantage of
providing exact, symbolic coefficients rather than imprecise
floating-point numbers.
However, aside from the limitation to exact solutions, such
infinite-precision symbolic methods do not scale as well to large
problems as the approximate, iterative methods used in linear
regression. To evaluate this scalability issue, we conducted
experiments comparing these methods within the \texttt{Mathematica}
and \texttt{Sympy} CAS, using the Fibonacci (\texttt{fib}) benchmark,
$22$ base functions (including powers of the golden ratio $\varphi$
and its conjugate $\overline{\varphi}$ to enable an exact solution),
and $\numtraining\in[10,100]$ data points.
It is worth noting that this example is intended solely to provide
orders of scale of computation time: it is small enough to allow the
use of exact symbolic methods and, for the same reason, employs
handpicked base functions that may not be available in practice.

\newdimen\abovecrulesep
\newdimen\belowcrulesep
\abovecrulesep=-0.3pt %
\belowcrulesep=0pt
\makeatletter
\patchcmd{\@@@cmidrule}{\aboverulesep}{\abovecrulesep}{}{}
\patchcmd{\@xcmidrule}{\belowrulesep}{\belowcrulesep}{}{}
\makeatother

\begin{table}[h!]
\centering
{\footnotesize
 \setlength{\extrarowheight}{2pt}
 \begin{tabular}{|| c || c | c | c | c || c ||}
 \cline{1-6}
 \multirow{2.25}{*}{Tool}
 &  \multicolumn{4}{c||}{$n$}
 & Output Solution \\
 \cmidrule(r{2.5pt}){2-5}
   & 10 & 20 & 34 & 100
   & (if applicable) \\
 \cline{1-6}
 \texttt{mlsolve(L)} &
   $\texttt{< 0.2s}$ &
   $\texttt{< 0.2s}$ &
   $\texttt{< 0.2s}$ &
   $\texttt{< 0.2s}$ &
   $\texttt{0.45}\cdot(\varphi^n-\overline{\varphi}^n)$\\
 \cline{1-6}
 \makecell[c]{\texttt{Mathematica}\\(symbolic)} &
   $\texttt{28.8s}$ &
   \makecell[c]{\phantom{.}\\[-0.9em]$\texttt{ O.M.}$\\[-2pt]\texttt{(> 30min)}} &
   $\texttt{T.O./O.M.}$ &
   $\texttt{T.O./O.M.}$ &
   \makecell[c]{\phantom{.}\\[-0.9em]$\frac{1}{\sqrt{5}}(\varphi^n-\overline{\varphi}^n),$\\{+ NullSpace}\\\phantom{.}\\[-0.9em]}\\
 \cline{1-6}
 \texttt{Sympy} &
   $\texttt{und.}$ &
   $\texttt{und.}$ &
   $\texttt{2.0s}$ &
   $\texttt{4.3s}$ &
   \makecell[c]{\phantom{.}\\[-0.9em]$\frac{\sqrt{5}}{5}\varphi^n-\frac{\sqrt{5}}{5}\overline{\varphi}^n$\\\phantom{.}\\[-0.9em]}\\
 \cline{1-6}
 \end{tabular}
}
 \caption{\label{table:symblineq-exp} Comparison of linear regression
   and symbolic linear equation solvers for coefficient search on
   the \texttt{fib} benchmark, with $22$ base functions and $n$ training
   points. Legend: underdetermined systems (\texttt{und.}), timeouts
   (\texttt{T.O.}), and out-of-memory errors (\texttt{O.M.}).}
\end{table}

For \texttt{Mathematica}, we use the \texttt{LinearSolve} function on
symbolic inputs, which returns one of the possible solutions when the
system is underdetermined (other functions can be used to compute the
full set of solutions).
For $n=10$ input points, we observe
that the solution it proposes (possibly chosen for its parsimony)
corresponds to the exact closed form for the \texttt{fib} benchmark,
although it obtains it at a $\texttt{>\,100}\times$ overhead compared
to $\texttt{mlsolve(L)}$.
In fact, this CAS obtains \texttt{O.M.}/\texttt{T.O.} errors for
$n\geq 20$, and is not able to reach the stage where the systems
become fully determined.
Investigating this issue shows that it is caused by an inappropriate
symbolic simplification strategy, where \texttt{Mathematica} conserves
large polynomial coefficients in $\qq[\sqrt{5}]$ during computation,
without simplifying them to $0$ early enough when possible, leading to
wasted computational resources (note that this only happens for
symbolic and not for \texttt{float}-based linear equation solving).

The CAS \texttt{Sympy} does not display this performance issue, and is
able to obtain the exact symbolic solution as soon as the system is
fully determined, although at a $\texttt{>\,10}\times$ overhead compared to
numerical iterative methods.

\texttt{mlsolve(L)} obtains a numerical approximation (with the
correct features) of the expected solution, spending less than $0.2s$
in regression for all chosen values of $n$, suggesting that this
time is mostly spent in bookkeeping tasks rather than the core of
regression itself.
Note that \texttt{mlsolve(L)}'s focus on sparse candidates allows to
obtain the correct solution even for underdetermined benchmarks.

Nonetheless, symbolic linear equation solving can be utilized in
certain cases to achieve exact symbolic regression, albeit at the
expense of time. As a direction for future work, it could be
integrated with our method and applied selectively when a solution is
deemed feasible, focusing on a subset of base functions identified by
Lasso and a subset of data points for efficiency.

\section{Related Work}
\label{sec:related-work}

\paragraph[Exact Recurrence Solvers]
{\textbf{Exact Recurrence Solvers}.}%

Centuries of work on recurrence equations have created a large body of
knowledge, whose full account is a matter of mathematical history
\citep{Dickson1919-NumberHistory-XVII}, with classical results such as
closed forms of \emph{C-recursive sequences}, i.e., of solutions of linear
recurrence equations with constant coefficients
\citep{Kauers2010-Tetrahedron, Petkovsek2013sketch}.

Despite important decision problems remaining open \citep{OuaknineSkolem}, the
field of symbolic computation now offers multiple algorithms to obtain solutions
and/or qualitative insights on various classes of recurrence equations. For
(monovariate) linear equations with polynomial coefficients, whose solutions are
named \emph{P-recursive sequences}, computing all their polynomial, rational and
hypergeometric solutions, whenever they exist, is a closed problem
\citep{Petkovsek92, AeqB, Abramov94}.%

Several of the algorithms available in the literature have been implemented in
popular Computer Algebra Systems such as \texttt{Sympy} \citep{sympy},
\texttt{Mathematica} \citep{mathematica-v13-2}, \texttt{Maple}
\citep{maple-heck} and \texttt{Matlab} \citep{matlab}.
These algorithms are built on insights coming from mathematical frameworks
including
\begin{itemize}
  \item \emph{Difference Algebra} \citep{Karr81, Bronstein00, ABPS21, Levin08},
    which considers wide classes of sequences via towers of
    $\Pi\Sigma^*$-extensions, and creates analogies between recurrence equations
    and differential equations,
  \item \emph{Finite Calculus} \citep{Gleich-2005-finite-calculus}, used in
    parts of \ciao's builtin solver, partly explaining its good
    results in the CAS category of our benchmarks,
  \item \emph{Operational Calculus} \citep{Berg67, kincaid2018}, and
  \item \emph{Generating Functions} \citep{Wilf94, Flajolet09},
\end{itemize}
which may be mixed with simple \emph{template-based methods}, e.g.,
finding polynomial solutions of degree $d$ by plugging such polynomial in the
equation and solving for coefficients.

Importantly, all of these techniques have in common the central role given
to the relation between a \emph{sequence} $f(n)$ and the \emph{shifted sequence}
$f(n-1)$, via, e.g., a shift operator $\sigma$, or multiplication by the formal
variable of a generating function.
As discussed before, it turns out that this simple observation highlights an
important obstacle to the application of CAS to program analysis via generalized
recurrence equations: these techniques tend to focus on
\emph{monovariate} recurrences built on \emph{finite differences}, i.e., on
recursive calls of shape $f(n-k)$ with $k$ constant, instead of more general
recursive calls $f(\phi(n))$.
Generalizations of these approaches exist, to handle multivariate ``partial
difference equations'', or ``$q$-difference equations'' with recursive calls
$f(q\cdot n)$ with $q\in\mathbb{C}$, $|q|<1$, but this is insufficient to deal
with recurrences arising in static cost analysis, which may contain inequations,
difficult recursive calls that cannot be discarded via change of variables,
piecewise definitions, or non-determinism.

CAS hence tend to focus on \emph{exact resolution} of a class of recurrence
equations that is quite different from those that arise in static cost analysis,
and do not provide bounds or approximations for recurrences they are unable to
solve.

In addition to classical CAS, we have tested \texttt{PURRS} \citep{BagnaraPZZ05},
which shares similarities with these solvers. \texttt{PURRS} is however more
aimed at automated complexity analysis, which is why it provides some support
for approximate resolution, as well as handling of some non-linear,
multivariate, and divide-and-conquer equations.

\paragraph[Recurrence Solving for Invariant Synthesis and Verification]
{\textbf{Recurrence Solving for Invariant Synthesis and Verification}.}

Another important line of work uses recurrence solving as a key ingredient in
generation and verification of program invariants, with further applications
such as loop summarization and termination analysis.
Much effort in this area has been put on improving \emph{complete} techniques
for restricted classes of programs, usually without recursion and built on
idealized numerical loops with abstracted conditionals and loop conditions.
These techniques can nevertheless be applied to more general programs, yielding
\emph{incomplete} approaches, via approximations and abstractions.

This line is well-illustrated by the work initiated in \citep{kovacs-phdthesis,
  kovacs08}, where \emph{all} polynomial invariants on program variables are
generated for a subclass of loops introduced in \citep{RodriguezKapur04}, using
recurrence-solving methods in addition to the ideal-theoretic approach of
\citep{RodriguezKapur07}. Recent work on the topic enabled inference of
polynomial invariants on combinations of program variables for a wider class of
programs with polynomial arithmetic \citep{kovacs18, kovacsSAS22}. A recommended
overview can be found in the first few sections of \citep{kovacs23}.

This approach is also key to the compositional recurrence analysis line of work
\citep{cra15, kincaid2017-short, kincaid2018, Breck-CHORA, kincaid2023},
implemented in various versions of \texttt{duet}, with a stronger focus on
abstraction-based techniques (e.g., the \emph{wedge} abstract domain), ability
to discover some non-polynomial invariants, and some (discontinued) support for
non-linear recursive programs, although the approach is still affected by
limited
accuracy in disjunctive reasoning.

Idealized numerical loops with precise conditionals are tackled by
\citep{Wang-PRS23}, tested in this paper under the name \texttt{PRS23}, which
builds upon the work developed in \texttt{VIAP} and its recurrence solver
\citep{VIAP17, VIAPRecSolver}. \texttt{PRS23} focuses on restricted
classes of loops, with ultimately periodic case application. Hence,
the problem of precise invariant generation, with fully-considered
branching and loop conditions, for extended classes of loops and
recursive programs, is still left largely open.

In addition, it may be noted that \emph{size} analysis, although typically
encountered in the context of static cost analysis, can sometimes be seen as a
form of numerical invariant synthesis, as illustrated by \citep{LommenGiesl23},
which exploits closed form computations on \emph{triangular
weakly non-linear loops}, %
presented, e.g., in \cite{FrohnSAS20}.

\paragraph[Cost Analysis via (Generalized) Recurrence Equations]
{\textbf{Cost Analysis via (Generalized) Recurrence Equations}.}%

Since the seminal work of \cite{Wegbreit75}, implemented in \textsc{Metric},
multiple authors have tackled the problem of cost analysis of programs (either
logic, functional, or imperative) by automatically setting up recurrence
equations, before solving them, using either generic CAS or specialized
solvers, whose necessity were quickly recognized.
Beyond \textsc{Metric}, applied to cost analysis of \texttt{Lisp} programs,
other important early work include \textsc{ACE} \citep{Le88-short}, and
\cite{Rosendahl89} in an abstract interpretation setting.

We refer the reader to previous publications in the \ciao line of work for
further context and details
\citep{granularity-short,caslog-short,low-bounds-ilps97-short,resource-iclp07-short,plai-resources-iclp14-short,gen-staticprofiling-iclp16-short,resource-verification-tplp18-shortest}.

We also include tools such as
\texttt{PUBS}\footnote{\url{https://costa.fdi.ucm.es/~costa/pubs/pubs.php}}
\citep{AlbertAGP11a-short, AlbertGM13} and
\texttt{Cofloco}\footnote{\url{https://github.com/aeflores/CoFloCo}}
\citep{montoya-phdthesis} in this category. These works emphasize the
shortcomings of using too simple recurrence relations, chosen as to fit the
limitations of available CAS solvers, and the necessity to consider
non-deterministic (in)equations, piecewise definitions, multiple variables,
possibly increasing variables, and to study non-monotonic behavior and
control-flow of the equations. They do so by introducing the vocabulary of
\emph{cost relations}, which may be seen as systems of non-deterministic
(in)equations, and proposing new coarse approximate resolution methods.

It may be noted that \texttt{duet}, mentioned above, also proposes an option for
resource bound analysis, as a specialization of its numerical analysis.
Additionally, recent work in the cost analyzer \texttt{KoAT} has given a greater
importance to recurrence solving, e.g., in \cite{LommenGiesl23}.

\paragraph[Other Approaches to Static Cost Analysis]
{\textbf{Other Approaches to Static Cost Analysis}.}%

Automatic static cost analysis of programs is an active field of research, and
many approaches have been proposed, using different abstractions than recurrence
relations.

For functional programs, type systems approaches have been studied. This
includes the concept of \emph{sized types} \citep{vh-03-short,
  vasconcelos-phdthesis}, but also the potential method implemented in
\texttt{RaML} \citep{jan-hoffman-phd, DBLP:journals/toplas/0002AH12}.

The version of \texttt{RaML} (1.5) tested in this paper is limited to potentials
(and hence size and cost bounds) defined by multivariate polynomials of bounded
degrees. One of the powerful insights of \texttt{RaML} is to represent
polynomials, not in the classical basis of monomials $x^k$, but in the binomial
basis $\binom{x}{k}$, leading to much simpler transformations of potentials when
type constructors are applied. Thanks to this idea, the problem of inference of
non-linear bounds can be reduced to a linear programming problem. A promising
extension of \texttt{RaML} to \emph{exponential} potentials has recently been
presented \citep{Kahn20}, using Stirling numbers of the second kind instead of
binomial coefficients, but, at the time of writing this paper, no implementation
is available to the best of the authors' knowledge.

Other important approaches include those implemented in \texttt{Loopus}, via
size-change graphs \citep{ZulegerSAS11} and difference constraints
\citep{Sinn17-loopus}, as well as those implemented in \texttt{AProve}
\citep{aprove-jar-2017}, \texttt{KoAT} \citep{GieslKoat22} and \texttt{LoAT}
\citep{loat-descr-22}, which translate programs to (extensions of) Integer
Transition Systems (ITS) and use, among other techniques, automated inference of
ranking functions, summarization, and alternate cost and size inference
\citep{Frohn-LPAR17}.

\paragraph[Dynamic Inference of Invariants/Recurrences]
{\textbf{Dynamic Inference of Invariants/Recurrences}.}%

Our approach is related to the line of work on dynamic invariant
analysis, which proposes to identify likely properties over variables
from observed program traces.
Pioneer work on this topic is exemplified by the tool
\texttt{Daikon}~\citep{ernst-phdthesis, daikon-tse}, which is able to
infer some linear relationships among a small number of explicit program
variables, as well as some template ``derived variables'', although it is
limited in expressivity and scalability.
Further work on dynamic invariant analysis made it possible to
discover invariants among the program variables that are
polynomial~\citep{nguyen2012:dyn-invariants-icse} and
tropical-linear~\citep{nguyen2014:disjunct-invariants-icse}, i.e., a
subclass of piecewise affine functions.
More recently, \citep{nguyen2022:invariants-tse} added symbolic
checking to check/remove spurious candidate invariants and obtain
counterexamples to help inference, in a dynamic, iterative guess and
check method, where the checking is performed on the program code.
Finally, making use of these techniques, an approach aimed at learning
asymptotic complexities, by dynamically inferring linear and
divide-and-conquer recurrences before extracting complexity orders from
them, is presented in~\citep{ishimwe2021:dynaplex-oopsla}.

While these works are directly related to ours, as they take advantage
of sample traces to infer invariants, there are several, important
differences from our work.
A key difference is that, %
instead of working directly on the program code, our method processes
recurrence relations, which may be seen as abstractions of the program
obtained by previous static analyses, and applies regression to
training sets obtained by evaluating the recurrences on input
``sizes'' instead of concrete data.\footnote{Note that such recurrence
relations can be seen as invariants on the cost of predicates, and are
useful in themselves for a number of applications, although for some
other applications we are interested in representing them as
closed-form functions.}
Hence, we apply dynamic inference techniques on already abstracted
programs: this allows complex invariants to be represented by simple
expressions, which are thus easier to discover dynamically.

Moreover, our approach differs from these works in the kind of
invariants that can be inferred, and in the regression techniques
being used.
The approach presented in \citep{nguyen2012:dyn-invariants-icse}
discovers polynomials relations (of bounded degree) between program
variables.
It uses an equation solving method to infer equality invariants --
which correspond to exact solutions in our context -- although their
work recovers some inequalities by other means.
Similarly to one instantiation of our guess method (but not
both), they generate templates that are affine combinations of a
predefined set of terms, which we call base functions.
However, unlike~\citep{nguyen2012:dyn-invariants-icse} we obtain
solutions that go beyond polynomials using both of our guessing
methods, as well as approximations. These approximations can be highly
beneficial in certain applications, as discussed in
Section~\ref{sec:intro} and further elaborated on in the following
section.

The Dynaplex~\citep{ishimwe2021:dynaplex-oopsla} approach
to dynamic complexity inference has different goals than ours: it aims
at asymptotic complexity instead of concrete cost functions, and, in
contrast to our approach, does not provide any soundness guarantees.
Dynaplex uses linear regression to find recurrences, but applies the
Master theorem and pattern matching to obtain closed-form expressions
representing asymptotic complexity bounds, unlike our approach, which
uses a different solving method and obtains concrete cost functions
(either exact or approximated), instead of complexity orders.

Nevertheless, it must be noted that all of these works are
complementary to ours, in the sense that they can be used as part of
our general framework. Indeed, we could represent recurrence relations
as programs (with input arguments representing input data sizes and
one output argument representing the cost) and apply, e.g.,
\citep{nguyen2012:dyn-invariants-icse} to find polynomial closed-forms
of the recurrences.
Similarly, we could apply the approach
in~\citep{nguyen2014:disjunct-invariants-icse}, which
is able to infer piecewise affine invariants using tropical polyhedra,
in order to extend/improve our domain splitting technique.

\section{Conclusions and Future Work}
\label{sec:conclusions}

We have developed a novel approach for solving or approximating
arbitrary, constrained recurrence relations.  It consists of a
\emph{guess} stage that uses machine learning techniques
to infer a candidate closed-form solution, and a \emph{check} stage
that combines an SMT-solver and a CAS to verify that such candidate is
actually a solution.
We have implemented a prototype and evaluated it within the context of
\ciaopp, a system for the analysis of logic programs (and other
languages via Horn clause tranformations).
The \emph{guesser} component of our approach is parametric w.r.t. the
machine learning technique used, and we have instantiated it with both
\emph{sparse (lasso) linear regression}
and \emph{symbolic regression} in our evaluation.

The experimental results are quite promising, showing that in general,
for the considered benchmark set, our approach outperforms
state-of-the-art cost analyzers and recurrence solvers. It also can
find closed-form solutions for recurrences that cannot be solved by
them. The results also show that our approach is reasonably efficient
and scalable.

Another interesting conclusion we can draw from the experiments is that
our machine learning-based solver could be used as a complement of
other (back-end) solvers in a combined higher-level solver, in order
to obtain further significant accuracy improvements. For example,
its combination with \ciaopp's builtin solver will potentially result
in a much more powerful solver, obtaining exact solutions for $88\%$ of the
benchmarks and accurate approximations for the rest of them, except
for one benchmark (which can only be solved by \texttt{PRS23}'s
solver).

Regarding the impact of this work on logic programming, note that an
improvement in a recurrence solver can potentially result in
arbitrarily large accuracy gains in cost analysis of (logic) programs.
Not being able to solve a recurrence can cause huge accuracy losses,
for instance, if such a recurrence corresponds to a predicate that is
deep in the control flow graph of the program, and such accuracy loss
is propagated up to the main predicate, inferring no useful
information at all.

The use
of regression techniques (with a randomly generated training set by
evaluating the recurrence to obtain the dependent value) does not
guarantee that a solution can always be found. Even if an exact
solution is found in the \emph{guess} stage, it is not always possible
to prove its correctness in the \emph{check} stage. Therefore, in this
sense, our approach is \emph{not complete}.
However, note that in the case where our approach does not obtain an
exact solution, the closed-form candidate inferred by the \emph{guess}
stage, together with its \emph{accuracy score}, can still be very
useful in some applications (e.g., granularity control in
parallel/distributed computing), where good approximations work well,
even though they are not upper/lower bounds of the exact solutions.

As a proof of concept, we have considered a particular deterministic
evaluation for constrained recurrence relations, and the verification
of the candidate solution is consistent with this evaluation. However,
it is possible to implement different evaluation semantics for the
recurrences, to support, e.g., non-deterministic or probabilistic
programs, adapting the verification stage accordingly.
Note that we need to require termination of the recurrence
evaluation as a precondition in verification of the obtained results.
This is partly due to the particular evaluation strategy of
recurrences that we are considering. In practice, non-terminating
recurrences can be discarded in the \emph{guess} stage, by setting a
timeout. Our approach can also be combined with a termination prover
in order to guarantee such a precondition.

As a future work, we plan to fully integrate our novel solver into the
\ciaopp system, combining it with its current set of back-end solvers
(referred to as \ciaopp's builtin solver in this paper) in order to
improve the static cost analysis. As commented above, the experimental
results encourage us to consider such a potentially powerful
combination, which, as an additional benefit, would allow \ciaopp
to avoid the use of external commercial solvers.

We also plan to further refine and improve our algorithms in several
directions. As already explained, the set $\basefuns$ of base
functions is currently fixed, user-provided. We plan to automatically
infer it by using different heuristics. We may perform an automatic
analysis of the recurrence we are solving, to extract some features
that allow us to select the terms that most likely are part of the
solution. For example, if the system of recurrences involves a
subproblem corresponding to a program with doubly nested loops, we can
select a quadratic term, and so on. Additionally, machine learning
techniques may be applied to learn a suitable set of base functions
from selected recurrence features (or the programs from which they
originate).  Another interesting line for future work is to extend our
solver to deal (directly) with non-deterministic recurrences.

Finally, we plan to use the counterexamples found by the
\emph{checker} component to provide feedback (to the \emph{guesser})
and help refine the search for better candidate solutions, such as by
splitting the recurrence
domains. Although our current domain splitting strategy already
provides good improvements, it is purely syntactic. We also plan to
develop more advanced strategies, e.g., by using a generalization of
model trees.

\bibliographystyle{tlplike}

\vfill

\appendix

\section{New Contributions w.r.t. our Previous Work}
\label{appendix:new-contributions}

\vspace{-0.1cm}

This work is an extended and revised version of our previous work
presented at ICLP 2023 as a Technical
Communication~\citep{ml-rec-solving-iclp2023}. The main additions and
improvements include:

\begin{itemize}

\item We report on a much more extensive experimental evaluation using
  a larger, representative set of increasingly complex benchmarks,
  and compare our approach with recurrence solving capabilities of
  state-of-the-art cost analyzers and CASs. In particular, we compare
  it with RaML, PUBS, Cofloco, KoAT, Duet, Loopus, PRS23, Sympy, PURRS
  and Mathematica.

\item Since our \emph{guess-and-check} approach is parametric in the
  regression technique used, in~\citep{ml-rec-solving-iclp2023} we
  instantiate it to (sparse) linear regression, but here we also
  instantiate it to symbolic regression, and compare the results
  obtained with both regression methods.

\item In Section~\ref{sec:description-approach-domain-splitting} we
  introduce a technique that processes multiple subdomains of the
  original recurrence separately, which we call \emph{domain
  splitting}. This strategy is orthogonal to the regression method
  used, and improves the results obtained, as our experimental
  evaluation shows.

\item In Section~\ref{sec:related-work} we include a more extensive
  discussion on related work.

\item In general, we have made some technical improvements, including
  a better formalization of recurrence relation solving.
  
\end{itemize}  

\pagebreak
\newpage

\section{Extended Experimental Results -- Full Outputs}
\label{appendix:full-output}

In Table~\ref{table:newexp1}, to provide a more comprehensive overview
within the constraints of available space, only a few symbols are used
to categorize the outputs of the tools. For the sake of completeness,
we include here full outputs of the tools in cases where approximate
solutions were obtained (symbols ``\approxbnd'' and ``\bad'').
We do not include exact outputs (symbol ``\good''), since they are simply
equivalent to those in Table~\ref{table:newexp2}. We neither include details and
distinctions in case where we obtained errors, trivial $+\infty$ bounds,
timeouts, or were deemed unsupported, as they are better suited for direct
discussions with tool authors.

\vspace{-2pt}
\begin{multicols}{2}

\noindent\textit{\textbf{\texttt{highdim1}}}
\setlist{nolistsep}
\begin{itemize}[leftmargin=*]\footnotesize
  \item \texttt{mlsolve(S)+domsplit}.\\
        $x_1 + 6.40410461112967\cdot x_3 + 8.31920033464576\cdot x_4 + 6.40410461112967\cdot x_5 + 8.31920033464576\cdot x_6 + 8.31920033464576\cdot x_7 + 8.31920033464576\cdot x_8 + 8.31920033464576\cdot x_9$ (case $\forall i, x_i>0$)
  \item \texttt{PUBS}.\\
        $11+10\cdot\max(0,x_1+x_2+x_3+x_4+x_5+x_6+x_7+x_8+x_9+x_{10})$
  \item \texttt{KoAT}.\\
        $x_1+2 x_2+3 x_3+4 x_4+5 x_5+6 x_6+7 x_7+8 x_8+9 x_9+10 x_{10}+66$
\end{itemize}
\textit{\textbf{\texttt{poly7}}}
\setlist{nolistsep}
\begin{itemize}[leftmargin=*]\footnotesize
  \item \texttt{mlsolve(L)}.\\
        $txyz + 0.02\cdot txy + 0.88\cdot tx + x^3z + z^4 - 0.01\cdot z^3 + 6.16$
  \item \texttt{mlsolve(L)+domsplit}.\\
        $txyz + 0.95\cdot tx + x^3z + z^4 - 0.01\cdot z^3 + 5.76$
\end{itemize}
\textit{\textbf{\texttt{highdim2}}}
\setlist{nolistsep}
\begin{itemize}[leftmargin=*]\footnotesize
  \item \texttt{mlsolve(S)+domsplit}.\\
        $x_1^2x_5x_6\cdot x_3!\cdot\max(x_5, x_9!)$ (case $x_1>0$)
  \item \texttt{Loopus15 (*)}.\\
        $x_1\max(x_1,x_2)\max(x_1,x_2,x_3)x_4x_5x_6x_7x_8x_9x_{10}$
\end{itemize}
\textit{\textbf{\texttt{loop\_tarjan}}}
\setlist{nolistsep}
\begin{itemize}[leftmargin=*]\footnotesize
  \item \texttt{KoAT} and \texttt{Loopus15 (*)}.\\
        $2n+1$
\end{itemize} 
\textit{\textbf{\texttt{enqdeq1}}}
\setlist{nolistsep}
\begin{itemize}[leftmargin=*]\footnotesize
  \item \texttt{Cofloco}.\\
        $4n$
  \item \texttt{Loopus15 (*)}.\\
        $n^2+2n$
\end{itemize}
\textit{\textbf{\texttt{enqdeq2}}}
\setlist{nolistsep}
\begin{itemize}[leftmargin=*]\footnotesize
  \item \texttt{mlsolve(L)}.\\
        $-0.53\cdot k + 0.06\cdot m n + 0.96\cdot m + 0.72\cdot n + 6.64$
  \item \texttt{mlsolve(L)+domsplit}.\\
        $-0.46\cdot k + 0.06\cdot m n + 0.9\cdot m + 0.95\cdot n + 5.21$
  \item \texttt{mlsolve(S)+domsplit}.\\
        $\max(n\mathord{+}m,\,1.7164959\mathord{\cdot}(n\mathord{+}m)\mathord{-}\linebreak\;\;\max(8.868849717670328,k))$
  \item \texttt{RaML}.\\
        $2n+m$
  \item \texttt{Duet-CRA23 (*)}.\\
        Raw\;{\scriptsize\texttt{max:302(max:302(max:302((param1:37 + (2 * param0:35)),
          param1:37), param0:35), 0)}}.\\
        Simplifies to $2n+m$
  \item \texttt{Cofloco}.\\
        Without disjunctive bounds,\\
        $\max(\max(2m,2\max(0,k)\mathord{+}\max(0,m\mathord{-}k)),\\\max(n\mathord{+}2m,\max(0,-n\mathord{+}m\mathord{-}k)\mathord{+}\max(0,2n\mathord{+}k)+\max(0,n\mathord{+}k))\mathord{+}n)$\\
        With disjunctive bounds, 7 cases, simplifies to\\
        {\scriptsize$\begin{cases}
          n &\text{if}\; m=0\\
          2n+2m &\text{if}\; m\geq1, m\leq n+k\\
          3n+m+k &\text{if}\; m\geq1, m\geq n+k+1\end{cases}$}
  \item \texttt{Loopus15 (*)}.\\
        $mn+m+n$
\end{itemize}
\textit{\textbf{\texttt{enqdeq3}}}
\setlist{nolistsep}
\begin{itemize}[leftmargin=*]\footnotesize
  \item \texttt{RaML} and \texttt{Duet-CRA23 (*)}.\\
        $8n$
  \item \texttt{Cofloco}.\\
        $9n+\max(0,4n-1)$, simplified to $13n-1$ when $n>0$.
  \item \texttt{Loopus15 (*)}.\\
        $5n+5n^2$
\end{itemize}
\textit{\textbf{\texttt{merge-sz}}}
\setlist{nolistsep}
\begin{itemize}[leftmargin=*]\footnotesize
  \item \texttt{PUBS}.\\
        $\max(0,x+y-1)+\max(\max(0,y),\max(0,x))$,\\
        simplifies to $x+y + \max(x,y)-1$.
  \item \texttt{KoAT}.\\
        $3x+3y$
  \item \texttt{Loopus15 (*)}.\\
        $2x+2y$
\end{itemize}
\textit{\textbf{\texttt{merge}}}
\setlist{nolistsep}
\begin{itemize}[leftmargin=*]\footnotesize
  \item \texttt{mlsolve(L)}.\\
        $x+y-1$
  \item \texttt{Ciao's builtin}, \texttt{RaML} and \texttt{Loopus15 (*)}.\\
        $x+y$
  \item \texttt{KoAT}.\\
        $3x+3y$
  \item \texttt{Duet-ICRA18}.\\
        Raw\;{\scriptsize\texttt{max:796(max:796(max:796(max:796((-1 + param0:57 + param1:60), 1),
                                  (-1 + param0:57 + param1:60)),
                          0),
                  (-1 + param0:57 + param1:60))}}.\\
        Simplifies to $\max(1,x+y-1)$
  \item \texttt{Duet-CHORA} and \texttt{Duet-CRA23 (*)}.\\
        $\max(0,x+y-1)$
\end{itemize}
\textit{\textbf{\texttt{open-zip}}}
\setlist{nolistsep}
\begin{itemize}[leftmargin=*]\footnotesize
  \item \texttt{RaML}, \texttt{PUBS}, \texttt{Duet-CHORA} and \texttt{Loopus15 (*)}.\\
        $x+y$
  \item \texttt{KoAT}.\\
        $x+2y$
  \item \texttt{Duet-ICRA18}.\\
        Raw\;{\scriptsize\texttt{max:1322(max:1322(max:1322(max:1322(param0:63, 0), 1), param1:66),
                   (-1 + param1:66 + param0:63))}}.\\
        Simplifies to $\max(1,x,y,x+y-1)$
\end{itemize}
\textit{\textbf{\texttt{s-max}}}
\setlist{nolistsep}
\begin{itemize}[leftmargin=*]\footnotesize
  \item \texttt{Ciao's builtin} and \texttt{PUBS}.\\
        $x+y+1$
  \item \texttt{KoAT}.\\
        $x+4y+1$
  \item \texttt{Duet-ICRA18}.\\
        Raw\;{\scriptsize\texttt{max:662(max:662(max:662((param1:52 + param0:49), (2 + param1:52)),
                          (1 + param1:52)),
                  param1:52)}}.\\
        Simplifies to $y+\max(2,x)$
  \item \texttt{Loopus15 (*)}.\\
        $x+2y$
\end{itemize} 
\ \\ \textit{\textbf{\texttt{s-max-1}}}
\setlist{nolistsep}
\begin{itemize}[leftmargin=*]\footnotesize
  \item \texttt{KoAT}.\\
        $3x+4y+1$
  \item \texttt{Duet-ICRA18}.\\
        Raw\;{\scriptsize\texttt{max:662(max:662(max:662(max:662(max:662((-1 + param1:52
                                                     + (2 * param0:49)),
                                                  (param1:52
                                                     + (2 * param0:49))),
                                          (3 + param1:52)),
                                  (2 + param1:52)),
                          param1:52),
                  (1 + param1:52))}}.\\
        Simplifies to $y+\max(2x,3)$
  \item \texttt{Duet-CRA23 (*)}.\\
        Raw\;{\scriptsize\texttt{max:136(max:136(max:136(param1:27, (1 + param1:27)),
                        (-1 + param1:27 + (2 * param0:25))),
                (param1:27 + (2 * param0:25)))}}.\\
        Simplifies to $y+\max(1,2x)$
  \item \texttt{Loopus15 (*)}.\\
        $3x+2y$
\end{itemize}
\textit{\textbf{\texttt{incr1}}}
\setlist{nolistsep}
\begin{itemize}[leftmargin=*]\footnotesize
  \item \texttt{mlsolve(L)}.\\
        $8.1 - 0.46\cdot x$
  \item \texttt{KoAT}.\\
        $x+10$
  \item \texttt{Duet-ICRA18}.\\
        Raw\;{\scriptsize\texttt{max:399(max:399(2,(11+(-1*param0:30))),1)}}.\\
        Simplifies to $\max(2,11-x)$
\end{itemize}
\textit{\textbf{\texttt{noisy\_strt1}}}
\setlist{nolistsep}
\begin{itemize}[leftmargin=*]\footnotesize
  \item \texttt{Ciao's builtin}, \texttt{RaML}, \texttt{PUBS}, \texttt{Duet-CRA23} and \texttt{Duet-CRA23 (*)}.\\
        $x$
  \item \texttt{Duet-CHORA}.\\
        Raw\;{\scriptsize\texttt{$\max(\max((-20 + x), 0), x)$}}.\\ 
        Simplifies to $x$
  \item \texttt{Duet-ICRA18}.\\
        Raw\;{\scriptsize\texttt{max:561(max:561(max:561(max:561(1, param0:30),
                                  min:560(19, param0:30)),
                          0),
                  (-20 + param0:30))}}.\\
        Simplifies to $\max(1,x)$
  \item \texttt{KoAT}.\\
        $2x$
  \item \texttt{Loopus15 (*)}.\\
       {\scriptsize ``\texttt{max(0,(x + -20)) assuming \{(>= x1 0)\}}''}
  \item \texttt{PURRS}.\\
        $x-20$ ``assuming $x\geq 20$''
  \item \texttt{Mathematica}.\\
        $0$ if $x=0$ or $x=20$, $x+cst$ otherwise
\end{itemize}
\textit{\textbf{\texttt{noisy\_strt2}}}
\setlist{nolistsep}
\begin{itemize}[leftmargin=*]\footnotesize
  \item \texttt{mlsolve(L)}, \texttt{Ciao's builtin}, \texttt{RaML}, \texttt{PUBS}, \texttt{Duet-CRA23} and \texttt{Duet-CRA23 (*)}.\\
        $x$
  \item {\scriptsize \texttt{mlsolve(L)+domsplit} and \texttt{mlsolve(S)+domsplit}.}\\
        $0$ if $x=0$ or $x=65536$, $x$ otherwise
  \item \texttt{KoAT}.\\
        $2x$
  \item \texttt{Duet-ICRA18}.\\
        Raw\;{\scriptsize\texttt{max:561(max:561(max:561(max:561(1, param0:30),
                                  min:560(65535, param0:30)),
                          0),
                  (-65536 + param0:30))}}.\\
        Simplifies to $\max(1,x)$
  \item \texttt{Loopus15 (*)}.\\
        {\scriptsize ``\texttt{max(0, (x + -65536) assuming \{(>= x1 0)\}}''}
  \item \texttt{PURRS}.\\
        $x-65536$ ``assuming $x\geq 65536$''
  \item \texttt{Mathematica}.\\
        $0$ if $x=0$ or $x=65536$, $x+\mathrm{cst}$ otherwise
\end{itemize}
\textit{\textbf{\texttt{multiphase1}}}
\setlist{nolistsep}
\begin{itemize}[leftmargin=*]\footnotesize
  \item \texttt{mlsolve(L)}.\\
        $-0.02\cdot i n - 0.6\cdot i + 1.83\cdot n + 0.59\cdot r - 6.49$
  \item \texttt{KoAT}.\\
        $nr+i+n+r$
  \item \texttt{Duet-ICRA18}.\\
        Raw\;{\scriptsize\texttt{max:372(max:372(max:372(0, param2:58), (param1:55 + param2:58)),
                  (param1:55 + (-1 * param0:52)))}}.\\
        Simplifies to $\max(n-i,n+r)$
  \item \texttt{Duet-CRA23} and \texttt{Duet-CRA23 (*)}.\\
        Raw\;{\scriptsize\texttt{max:201(max:201(max:201((param1:37 + -param0:35),
                                min:200(param2:39, -param0:35)),
                        (param2:39 + param1:37)),
                0)}}.\\
        Simplifies to $\max(n-i,n+r)$
  \item \texttt{Loopus15 (*)}.\\
        $r+nr+\max(0,n-i)$
\end{itemize}
\textit{\textbf{\texttt{lba\_ex\_viap}}}
\setlist{nolistsep}
\begin{itemize}[leftmargin=*]\footnotesize
  \item \texttt{mlsolve(L)+domsplit}.\\
        $.49\cdot(c-x-y)+.33$ if $x+y<c$, $0$ if $x+y\geq c$.
  \item \texttt{mlsolve(S)+domsplit}.\\
        $\lceil(c-(x+y))\cdot 0.473..\rceil $ if $x+y<c$, $0$ if $x+y\geq c$.
  \item \texttt{PUBS}, \texttt{Duet-ICRA18}, \texttt{Duet-CHORA}, \texttt{Duet-CRA23} and \texttt{Duet-CRA23 (*)}.\\
        $\max(0,c/2-x/2-y/2+1/2)$
  \item \texttt{Cofloco}.\\
        $-x/2-y/2+c/2$ if $x+y<c$, $0$ if $x+y\geq c$.
  \item \texttt{KoAT}.\\
        $x+y+c$
  \item \texttt{Loopus15 (*)}.\\
        $\max(0,c-x-y)$
\end{itemize}
\textit{\textbf{\texttt{nested}}}
\setlist{nolistsep}
\begin{itemize}[leftmargin=*]\footnotesize
  \item \texttt{Duet-ICRA18}.\\
        $\max(1,x)$
\end{itemize}
\textit{\textbf{\texttt{nested\_case}}}
\setlist{nolistsep}
\begin{itemize}[leftmargin=*]\footnotesize
  \item \texttt{mlsolve(L)}.\\
        $0.5\cdot x\cdot(x + 1)$
\end{itemize}
\textit{\textbf{\texttt{nested\_div}}}
\setlist{nolistsep}
\begin{itemize}[leftmargin=*]\footnotesize
  \item \texttt{mlsolve(L)}.\\
        $0.14\cdot\lceil\log_2(n)\rceil + 1.62$
  \item \texttt{mlsolve(L)+domsplit}.\\
        $0.24\cdot\lceil\log_2(n)\rceil + 0.12\cdot\lfloor\log_2(n)\rfloor + 1.7$ if $x>0$,
        $0$ if $x=0$.
  \item \texttt{RaML}.\\
        $x$
\end{itemize}
\textit{\textbf{\texttt{mccarthy91}}}
\setlist{nolistsep}
\begin{itemize}[leftmargin=*]\footnotesize
  \item \texttt{mlsolve(L)+domsplit}.\\
        $x-10$ if $x\geq 101$, \texttt{ERROR} if $x\leq 100$.
  \item \texttt{Cofloco}.\\
        $x-10$ if $x\geq 101$, $\inf$ if $x\leq 100$.
\end{itemize}
\textit{\textbf{\texttt{golomb}}}
\setlist{nolistsep}
\begin{itemize}[leftmargin=*]\footnotesize
  \item \texttt{mlsolve(L)}.\\
        $1.38\sqrt{n} + 0.06\cdot\lceil\log_2(n)\rceil - 0.16$
  \item \texttt{mlsolve(L)+domsplit}.\\
        $1.44\sqrt{n}-0.15$ if $x>0$,
        $1$ if $x=0$
  \item \texttt{mlsolve(S)+domsplit}.\\
        $\lceil x^{0.5857451} \rceil$
  \item \texttt{RaML}.\\
        $1+1.5x+.5x^2$
\end{itemize}
\textit{\textbf{\texttt{div}}}
\setlist{nolistsep}
\begin{itemize}[leftmargin=*]\footnotesize
  \item \texttt{PUBS} and \texttt{Duet-CHORA}.\\
        $x$
  \item \texttt{Cofloco} and \texttt{Loopus15 (*)}.\\
        $\max(0,x-y+1)$
  \item \texttt{KoAT}.\\
        $x+y+1$
  \item \texttt{Duet-ICRA18}.\\
        $\max(1,2+x-2y)$
  \item \texttt{Mathematica}.\\
        $x/y$ (without floor, if hinted that y is a constant)
\end{itemize}
\textit{\textbf{\texttt{sum-osc}}}
\setlist{nolistsep}
\begin{itemize}[leftmargin=*]\footnotesize
  \item \texttt{mlsolve(L)}.\\
        $.5\cdot y^2+1.5\cdot y+x$
  \item \texttt{mlsolve(S)+domsplit}.\\
        {\scriptsize$\begin{cases}
          \big\lfloor x - 0.9866641 + \frac{(y + 0.9866641)^3}{2y}\rfloor\\\qquad\text{if}\; x>0,y>0\\
          \lfloor 0.886569412557\cdot(0.756346845291\cdot y + 1)^2 \rfloor\\\qquad\text{if}\; x=0, y>0\\
          1\\\qquad\text{if}\; y=0\\\end{cases}$}
  \item \texttt{RaML}.\\
        $1+.5\cdot y^2+1.5\cdot y+x$
  \item \texttt{PUBS}.\\
        $1+\max(0,x+2y-1)\cdot\max(1,\max(0,y))$\\
        Simplifies to $1+xy+2y^2-y$
  \item \texttt{Cofloco}.\\
        $\begin{cases}
          \max(x+2y-1,(x+2y-1)*y)+1 &\text{if}\; y>0\\
          1 &\text{if}\; y=0\\\end{cases}$\\
        Simplifies to $1$ if $y=0$, $1+xy+2y^2-y$ if $y>0$.
  \item \texttt{KoAT}.\\
        $1+x+y+2y^2$
\end{itemize}
\textit{\textbf{\texttt{bin\_search}}}
\setlist{nolistsep}
\begin{itemize}[leftmargin=*]\footnotesize
  \item \texttt{mlsolve(L)}.\\
        $1+\lfloor\log_2(x)\rfloor$ (for all $x\geq 0$)
  \item \texttt{RaML}.\\
        $2x$
  \item \texttt{PUBS}.\\
        $\log_2(1+\max(0,2x-1))$\\
        Simplifies to $1$ if $x=0$, $1+\log_2(x)$ if $x>0$
  \item {\scriptsize \texttt{Duet-CHORA}, \texttt{Duet-CRA23 (*)} and \texttt{Loopus15 (*)}}.\\
        $x$
  \item \texttt{PURRS}.\\
        $1+\log_2(x)$ (ub), after overapproximating floor.
  \item \texttt{Mathematica}.\\
        $\log_2(x)+\mathrm{cst}$, after overapproximating floor and some rewriting.
\end{itemize}
\textit{\textbf{\texttt{qsort\_best}}}
\setlist{nolistsep}
\begin{itemize}[leftmargin=*]\footnotesize
  \item \texttt{mlsolve(L)}.\\
        $0.05\cdot x^2 + 0.1\cdot x \lceil\log_2(x)\rceil + 3.43\cdot x - 0.96\cdot \lceil\log_2(x)\rceil + 0.27\cdot \lfloor\sqrt{x}\rfloor - 0.77$
  \item \texttt{mlsolve(L)+domsplit}.\\
        $0.31\cdot x \lceil\log_2(x)\rceil + 3.8\cdot x - 2.64\cdot \lceil\log_2(x)\rceil + 0.33$ if $x>0$,
        $1$ if $x=0$.
  \item \texttt{mlsolve(S)+domsplit}.\\
        $\lceil 1.9746461095749\cdot (x + 0.179665875875)^{1.286233} \rceil$ \\if $x>0$,
        $1$ if $x=0$.
  \item \texttt{mlsolve(L)} (with $\bndtraining=100$).\\
        $0.33\cdot x \lfloor\log_2(x)\rfloor + 5.52\cdot x - 15.35\cdot \lfloor\log_2(x)\rfloor + 15.07$
  \item \texttt{mlsolve(L)+domsplit} (with $\bndtraining=100$).\\
        $0.34\cdot x \lceil\log_2(x)\rceil + 4.99\cdot x - 11.58\cdot \lceil\log_2(x)\rceil + 13.14$ if $x>0$,
        $1$ if $x=0$.
  \item \texttt{mlsolve(S)+domsplit} (with $\bndtraining=100$).\\
        $1.0148813\cdot x \log_2(x + 4.7831106)$ \\if $x>0$,
        $1$ if $x=0$.
  \item \texttt{RaML}.\\
        $1 + 4 x + 1/3 x^2$
  \item \texttt{PUBS}.\\
        $(\max(0,2x-1)+1-1)\cdot \max(0,x)$\\
        Simplifies to $2x^2-x$
  \item \texttt{Duet-CHORA}.\\
        $1+x\cdot 2^x$
  \item \texttt{PURRS}.\\
        $x\cdot\log_2(x)+3x$ (ub), after overapproximating floor/ceil and some rewriting.
  \item \texttt{Mathematica}.\\
        $x\cdot\log_2(x) + \mathrm{cst}\cdot x$, after overapproximating floor/ceil and some rewriting.
\end{itemize}
\textit{\textbf{\texttt{prs23\_1}}}
\setlist{nolistsep}
\begin{itemize}[leftmargin=*]\footnotesize
  \item \texttt{Duet-CHORA}.\\
        $f(n)\leq \max(1,1+499n)$,
        $g(n)\leq \max(1,3n)$
  \item \texttt{Duet-CRA23} and \texttt{Duet-CRA23 (*)} .\\
        Raw\;{\scriptsize$f(n)\leq$ \texttt{max:212(min:211(min:211((1 + (499 * param0:15)),
          998), pow:25(2, param0:15)),
          min:211(min:211(min:211(min:211(min:211(min:211( min:211(min:211(
          (-498 + ( 499 * param0:15)), (1 + ( 499 * log:26( 2, 998)))), (1 + (
          499 * log:26( 2, ( 1 + ( 499 * log:26( 2, 998))))))), (1 + ( 499 *
          log:26( 2, ( -498 + ( 499 * param0:15)))))), (1 + (499 * log:26( 2, (1
          + ( 499 * log:26( 2, ( -498 + ( 499 * param0:15))))))))), (-498 + (499
          * log:26(2, pow:25( 2, param0:15))))), 998), (1/2 * pow:25(2,
          param0:15))), (1/2 * pow:25(2, (1/2 * pow:25(2, param0:15))))))}},\\
        {\scriptsize$g(n)\mathord{\leq}$\,\texttt{max:212(1, min:211((-2+(3 *
          param0:15)), (1+(3 * param0:15) + (-3 * log:26(2, 500)))))}}.\\
        $f$ simplifies to $2^n$ if $2\leq n\leq 9$, $998$ if $n\geq 10$.\\
        $g$ simplifies to $\max(1,1+3n-3\log_2(500))\approx\max(1,3n-25.90...)$.
  \item \texttt{Loopus15 (*)}.\\
        $f(n) \leq 998$ and $g(n) \leq 1+3n$
\end{itemize}
\textit{\textbf{\texttt{exp2}}}
\setlist{nolistsep}
\begin{itemize}[leftmargin=*]\footnotesize
  \item \texttt{mlsolve(S)+domsplit}.\\
        $4\cdot 2^x$ (first candidate),
        $\lfloor 4\cdot 2^x-.027...\cdot x\rfloor$ (second candidate)
\end{itemize}
\textit{\textbf{\texttt{exp3}}}
\setlist{nolistsep}
\begin{itemize}[leftmargin=*]\footnotesize
  \item \texttt{mlsolve(L)}.\\
        $24879911.68\cdot x^2y^2 - 382407544.6\cdot x^2y + 910687706.81\cdot x^2 - 426104965.68\cdot xy^2 + 6615353066.27\cdot xy - 16347484322.02\cdot x + 1437670074.84\cdot y^2 - 22905346457.85\cdot y + 64530269113.55$
  \item \texttt{mlsolve(L)+domsplit}.\\
        $7007804.6\cdot x^2y^2 - 77376230.79\cdot x^2y + 49836049.42\cdot x^2 - 87385060.39\cdot xy^2 + 867536772.6\cdot xy + 70890458.63\cdot x + 178082073.72\cdot y^2 - 1539360337.12\cdot y - 1728336336.08$
\end{itemize}
\textit{\textbf{\texttt{fib}}}
\setlist{nolistsep}
\begin{itemize}[leftmargin=*]\footnotesize
  \item \texttt{mlsolve(L)}.\\
        $69.13\cdot x^2 + 10.73\cdot x\cdot \lceil\log_2(x)\rceil - 133.88\cdot x\cdot \lfloor\log_2(x) - 1079.01\cdot x + 553.65\cdot \lceil\log_2(x)\rceil - 135.4\cdot \lfloor\sqrt{x}\rfloor + 1549.33\cdot \lfloor\log_2(x) + 729.74$
  \item \texttt{mlsolve(L)+domsplit}.\\
        $0$ if $x=0$, $1$ if $x=1$,
        $107.11\cdot x^2 - 106.74\cdot x\cdot \lceil\log_2(x)\rceil - 239.41\cdot x\cdot \lfloor\log_2(x) - 1133.71\cdot x + 1896.18\cdot \lceil\log_2(x)\rceil - 706.19\cdot \lfloor\sqrt{x}\rfloor + 2708.6\cdot \lfloor\log_2(x)\rfloor – 1859.08$ if $x>1$.
  \item \texttt{mlsolve(S)+domsplit}.\\
        $0$ if $x=0$, $1$ if $x=1$,
        $0.453497199508044\cdot 1.61690120042049^x – 0.239336049422447$ if $x>1$.
  \item \texttt{PUBS}.\\
        $2^{\max(0,x-1)}$
  \item \texttt{Duet-CHORA}.\\
        $2^x$
\end{itemize}
\textit{\textbf{\texttt{harmonic}}}
\setlist{nolistsep}
\begin{itemize}[leftmargin=*]\footnotesize
  \item \texttt{mlsolve(L)}.\\
        $0.68\cdot\sqrt{x} + 0.21$
  \item \texttt{mlsolve(L)+domsplit}.\\
        $0$ if $x=0$,
        $0.82\cdot\sqrt{x} - 0.15\cdot\lceil\log_2(x)\rceil + 0.15$ if $x>0$
  \item \texttt{mlsolve(S)+domsplit}.\\
        $0$ if $x=0$,
        $\lfloor\log_2(0.58528787\cdot x + 1.9952444)\rfloor$ if $x>0$
\end{itemize} 
\textit{\textbf{\texttt{fact}}}
\setlist{nolistsep}
\begin{itemize}[leftmargin=*]\footnotesize
  \item \texttt{mlsolve(L)}.\\
        $x! + 16.0$
  \item \texttt{mlsolve(L)+domsplit}.\\
        $0.02\cdot 2^x + x! + 32.0$
  \item \texttt{Cofloco}.\\
        $1$
\end{itemize}
\textit{\textbf{\texttt{cas\_st1}}}
\setlist{nolistsep}
\begin{itemize}[leftmargin=*]\footnotesize
  \item \texttt{mlsolve(L)}.\\
        $176380.52\cdot 2^x\cdot x - 1246486.37\cdot 2^x - 346.58\cdot 5^x + 140930.04\cdot x^2 - 22021.02\cdot x\cdot \lceil\log_2(x)\rceil + 1019395.76\cdot x\cdot \lfloor\log_2(x)\rfloor + 150189.16\cdot x + 405206.28\cdot \lceil\log_2(x)\rceil - 1598623.07\cdot \lfloor\sqrt{x}\rfloor - 1533107.27\cdot \lfloor\log_2(x)\rfloor + 1799.42\cdot x! + 3447651.05$
  \item \texttt{mlsolve(L)+domsplit}.\\
        $1$ if $x=0$,
$176380.52\cdot 2^x\cdot x - 1246486.37\cdot 2^x - 346.58\cdot 5^x + 140930.04\cdot x^2 - 22021.02\cdot x\cdot \lceil\log_2(x)\rceil + 1019395.76\cdot x\cdot \lfloor\log_2(x)\rfloor + 150189.16\cdot x + 405206.28\cdot \lceil\log_2(x)\rceil - 1598623.07\cdot \lfloor\sqrt{x}\rfloor - 1533107.27\cdot \lfloor\log_2(x)\rfloor + 1799.42\cdot x! + 3447651.05$ if $x>0$
  \item \texttt{Cofloco}.\\
        $1$ if $x=0$, $2$ if $x>0$
\end{itemize}
\textit{\textbf{\texttt{cas\_st2}}}
\setlist{nolistsep}
\begin{itemize}[leftmargin=*]\footnotesize
  \item \texttt{mlsolve(L)}.\\
        $1532499438.63\cdot 2^x\cdot x - 10961243871.35\cdot 2^x - 2533831.22\cdot 5^x + 1109062299.38\cdot x^2 - 409557670.88\cdot x\cdot \lceil\log_2(x)\rceil + 10970271676.25\cdot x\cdot \lfloor\log_2(x)\rfloor + 774384179.03\cdot x + 3890181725.0\cdot \lceil\log_2(x)\rceil - 18733069879.29\cdot \lfloor\sqrt{x}\rfloor - 16344125381.56\cdot \lfloor\log_2(x)\rfloor + 9112306.73\cdot x! + 35710669116.92$
  \item \texttt{mlsolve(L)+domsplit}.\\
        $1$ if $x=0$,
        $1532499438.25\cdot 2^x\cdot x - 10961243867.81\cdot 2^x - 2533831.22\cdot 5^x + 1109062296.0\cdot x^2 - 409557670.74\cdot x\cdot \lceil\log_2(x)\rceil + 10970271677.52\cdot x\cdot \lfloor\log_2(x)\rfloor + 774384185.44\cdot x + 3890181726.66\cdot \lceil\log_2(x)\rceil - 18733069876.69\cdot \lfloor\sqrt{x}\rfloor - 16344125386.41\cdot \lfloor\log_2(x)\rfloor + 9112306.73\cdot x! + 35710669102.86$ if $x>0$.
\end{itemize}
\textit{\textbf{\texttt{cas\_st3}}}
\setlist{nolistsep}
\begin{itemize}[leftmargin=*]\footnotesize
  \item \texttt{mlsolve(L)}.\\
        $-0.04\cdot 2^x + 2.72\cdot x! + 320.0$
  \item \texttt{mlsolve(L)+domsplit}.\\
        $1$ if $x=0$,
        $0.01\cdot 2^x\cdot x - 0.13\cdot 2^x + 2.72\cdot x! + 320.0$ if $x>0$.
  \item \texttt{mlsolve(S)+domsplit}.\\
        $x! \cdot  2.71828$
\end{itemize}
\textit{\textbf{\texttt{cas\_st4}}}
\setlist{nolistsep}
\begin{itemize}[leftmargin=*]\footnotesize
  \item \texttt{mlsolve(L)}.\\
        $0.69\cdot 2^x - 0.02\cdot x^2 - 0.04\cdot x\cdot \lceil\log_2(x)\rceil + 0.13\cdot x\cdot \lfloor\log_2(x)\rfloor + 0.04\cdot x + 0.51\cdot \lceil\log_2(x)\rceil - 0.14\cdot \lfloor\sqrt{x}\rfloor - 1.01\cdot \lfloor\log_2(x)\rfloor – 0.35$
  \item \texttt{mlsolve(L)+domsplit}.\\
        $0$ if $x=0$,
        $0.69\cdot 2^x - 0.02\cdot x^2 - 0.03\cdot x\cdot \lceil\log_2(x)\rceil + 0.13\cdot x\cdot \lfloor\log_2(x)\rfloor + 0.08\cdot x + 0.39\cdot \lceil\log_2(x)\rceil - 0.17\cdot \lfloor\sqrt{x}\rfloor - 0.99\cdot \lfloor\log_2(x)\rfloor – 0.31$ if $x>0$.
  \item \texttt{mlsolve(S)+domsplit}.\\
        $2^x \cdot  0.69315$
\end{itemize}
\textit{\textbf{\texttt{cas\_st5}}}
\setlist{nolistsep}
\begin{itemize}[leftmargin=*]\footnotesize
  \item \texttt{mlsolve(L)}.\\
        $100563.63\cdot 2^x\cdot x - 710686.18\cdot 2^x - 197.6\cdot 5^x + 80351.45\cdot x^2 - 12555.04\cdot x\cdot \lceil\log_2(x)\rceil + 581209.77\cdot x\cdot \lfloor\log_2(x)\rfloor + 85630.47\cdot x + 231028.75\cdot \lceil\log_2(x)\rceil - 911456.69\cdot \lfloor\sqrt{x}\rfloor - 874103.34\cdot \lfloor\log_2(x)\rfloor + 1025.94\cdot x! + 1965682.95$
  \item \texttt{mlsolve(L)+domsplit}.\\
        $0$ if $x=0$,
        $100563.63\cdot 2^x\cdot x - 710686.18\cdot 2^x - 197.6\cdot 5^x + 80351.45\cdot x^2 - 12555.04\cdot x\cdot \lceil\log_2(x)\rceil + 581209.77\cdot x\cdot \lfloor\log_2(x)\rfloor + 85630.47\cdot x + 231028.75\cdot \lceil\log_2(x)\rceil - 911456.69\cdot \lfloor\sqrt{x}\rfloor - 874103.34\cdot \lfloor\log_2(x)\rfloor + 1025.94\cdot x! + 1965682.95$ if $x>0$.
  \item \texttt{mlsolve(S)+domsplit}.\\
        $2^x\cdot x! \cdot  0.5701515$
\end{itemize}

\end{multicols}

\end{document}